\documentclass[letter]{aa} 
\usepackage{graphicx}
\usepackage{adjustbox}
\usepackage{txfonts}
\usepackage{xcolor}

\begin{document}

   \title{Asteroseismology of the multiple stellar populations in the Globular Cluster M4}

   \subtitle{}

   \author{M. Tailo
          \inst{1}
          \and
          E. Corsaro
          \inst{2}
          \and
          A. Miglio
          \inst{1,5, 6}
          \and
          J. Montalb\'{a}n
          \inst{1,6}
          \and
          K. Brogaard
          \inst{3}
          \and
          A. P. Milone
          \inst{4}
          \and
          A. Stokholm
          \inst{1,3,5}
          \and
          G. Casali
          \inst{1,5}
          \and
          A. Bragaglia
          \inst{5}
          }

   \institute{
            Dipartimento di Fisica e Astronomia Augusto Righi, Università degli Studi di Bologna, Via Gobetti 93/2, I-40129 Bologna, Italy
            \email{marco.tailo@unibo.it,mrctailo@gmail.com}
         \and
             INAF – Osservatorio Astrofisico di Catania, Via S. Sofia, 78, 95123 Catania, Italy
        \and
             Stellar Astrophysics Centre, Department of Physics and Astronomy, Aarhus University, Ny Munkegade 120, DK-8000 Aarhus C, Denmark
        \and
             Dipartimento di Fisica e Astronomia “Galileo Galilei”, Univ. di Padova, Vicolo dell’Osservatorio 3, Padova, IT-35122
        \and 
            INAF-Osservatorio di Astrofisica e Scienza dello Spazio di Bologna, via Gobetti 93/3, I-40129 Bologna, Italy
        \and
            School of Physics and Astronomy, University of Birmingham, Birmingham B15 2TT, United Kingdom
            }
   \date{Received September XXXX; accepted March YYYY}


  \abstract{We present a new asteroseismic analysis of the stars in the Globular Cluster (GC) M4 based on the data collected by the K2 mission. We report the detection of solar-like oscillation in 37 stars, 32 red giant branch (RGB) and 6 red horizontal branch (rHB) stars, the largest sample for this kind of study in GC up to date. Combining information from asteroseismology and multi-band photometry we estimate both the masses and the radii of our targets. Our estimates are in agreement with independent sources, serving as a crucial verification of asteroseismology in the low metallicity regime.\\
  As M4 is an old GC, it hosts multiple stellar populations differing in light-element abundances and in helium mass fraction. 
  This generates a mass difference between the populations along the RGB, which in the case of M4 is estimated to be $0.017 M_\odot$. With this wealth of information we can assign population membership and estimate the average mass of the stellar populations, but the current uncertainties do not allow us to resolve this mass difference. The population membership and the seismic data of RGB and HB stars, allow us, however,  to assess the integrated mass loss along the RGB of the first generation stars in the cluster.  We obtain $\rm \Delta M=0.227 \pm0.028 M_\odot$, in good agreement with independent estimates. Finally, we observe the presence of a statistically significant mass-temperature gradient in the rHB stars. This represents the first direct, model-independent observation of the colour-temperature-mass correlation predicted by the theory.}


   \maketitle
%

\section{Introduction}

The search for solar like oscillations in GC is difficult. Among the Galactic Globular Cluster (GC), M4 is the closest and the only one currently accessible to asteroseismic inference. Early attempts with ground based instruments  by \cite{Stello09} reported hints of detection, but the low signal to noise ratio and the complexity of the spectra made the detection unambiguous. Similarly for another metal poor GCs, NGC\,6397, \citet[][]{Frandsen2007} tried to detect solar like oscillation but, also in this case, their results were uncertain due to low signal to noise ratio.

The NASA K2 mission \citep{howell_2014} observed M4 during its second observational campaign, providing $\sim80$~d of nearly continuous, high-precision photometric monitoring. Using K2 data, \cite{Miglio_2016} reported clear detection of solar-like oscillations in K giants in M4, obtaining estimates of mass, radius and age compatible with estimates from other methods such as studies of eclipsing binaries \citep[e.g. by][]{Kaluzny2013}, providing a crucial test of asteroseismology in the low-metallicity domain \citep[{[}Fe/H{]}=$-1.1$,][]{marino_2008,carretta2013}. However, given the limited sample of stars with detected oscillations (8 targets),  \cite{Miglio_2016} could not address key questions about M4, like providing a robust estimate of the mass loss along the red giant branch (RGB) or investigating the features of the horizontal branch (HB) stars.  
  
The crowding of a GC field and the drift of the spacecraft make the detection of solar-like oscillations challenging. \citet{Wallace19} developed a procedure to mitigate these issues and extracted more than 4000 clean, high-quality time-series of M4 stars, which are publicly available. In their analysis, Wallace and collaborators identified 55 stars -- the majority on the RGB -- showing potential evidence for solar-like oscillations.
These high quality observations can help solve some of the questions left open by Miglio and collaborators' work. 

Being an old GC, M4 hosts multiple populations \citep{renzini_2015,bastian_2018,Gratton_2019} with different light-elements abundance patterns and helium enhancement. In the case of M4, only two major groups are present: (i) a first generation of stars with light-elements abundances compatible with field stars of similar metallicity (e.g high [O/Fe] and low [C+N/Fe] and [Na/Fe], with the standard helium abundance for its metallicity), and (ii) a second generation of stars with different chemical patterns e.g. lower [O/Fe] and higher [C+N/Fe] and [Na/Fe] with a helium enhancement of 0.013 \citep{marino_2008,marino_2011,tailo_2019a}. 

In this paper, we extend the work presented in \citet{Miglio_2016} to the candidates identified in \cite{Wallace19} with four objectives: i) infer masses and radii of the stars in M4 and compare them with independent measurements; ii) test the asteroseismic properties of metal poor stars; iii) analyse the multiple populations in M4 \citep[see][]{tailo_2019a} with asteroseismology; iv) estimate the integrated mass loss along the RGB, by estimating the stellar parameters of red HB stars.

\section{Data sets and observational constraints}
\label{sec:obs_datasets}

\begin{figure*}
    \centering
    \includegraphics[width=0.93\columnwidth]{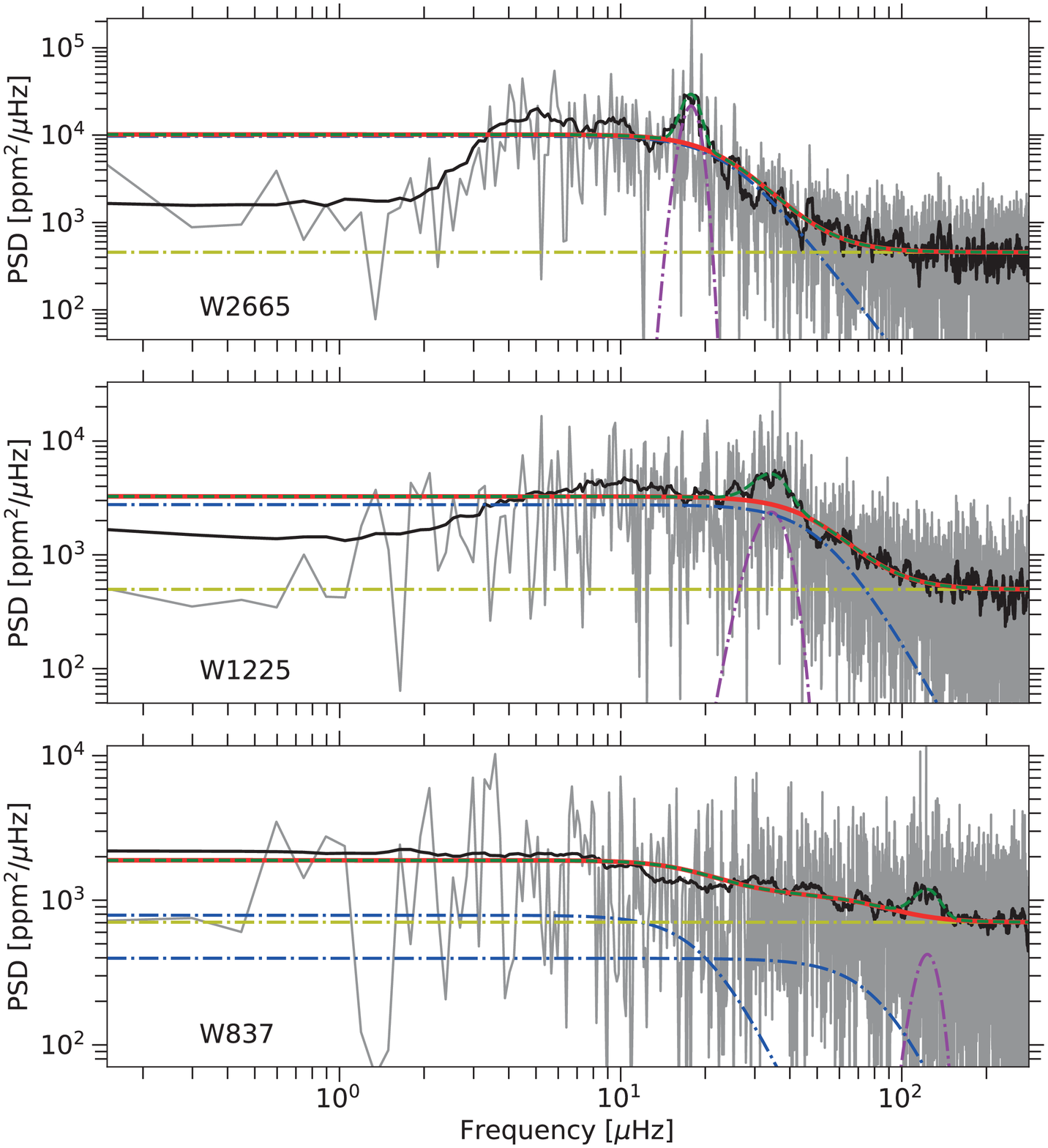}
    \includegraphics[width=\columnwidth]{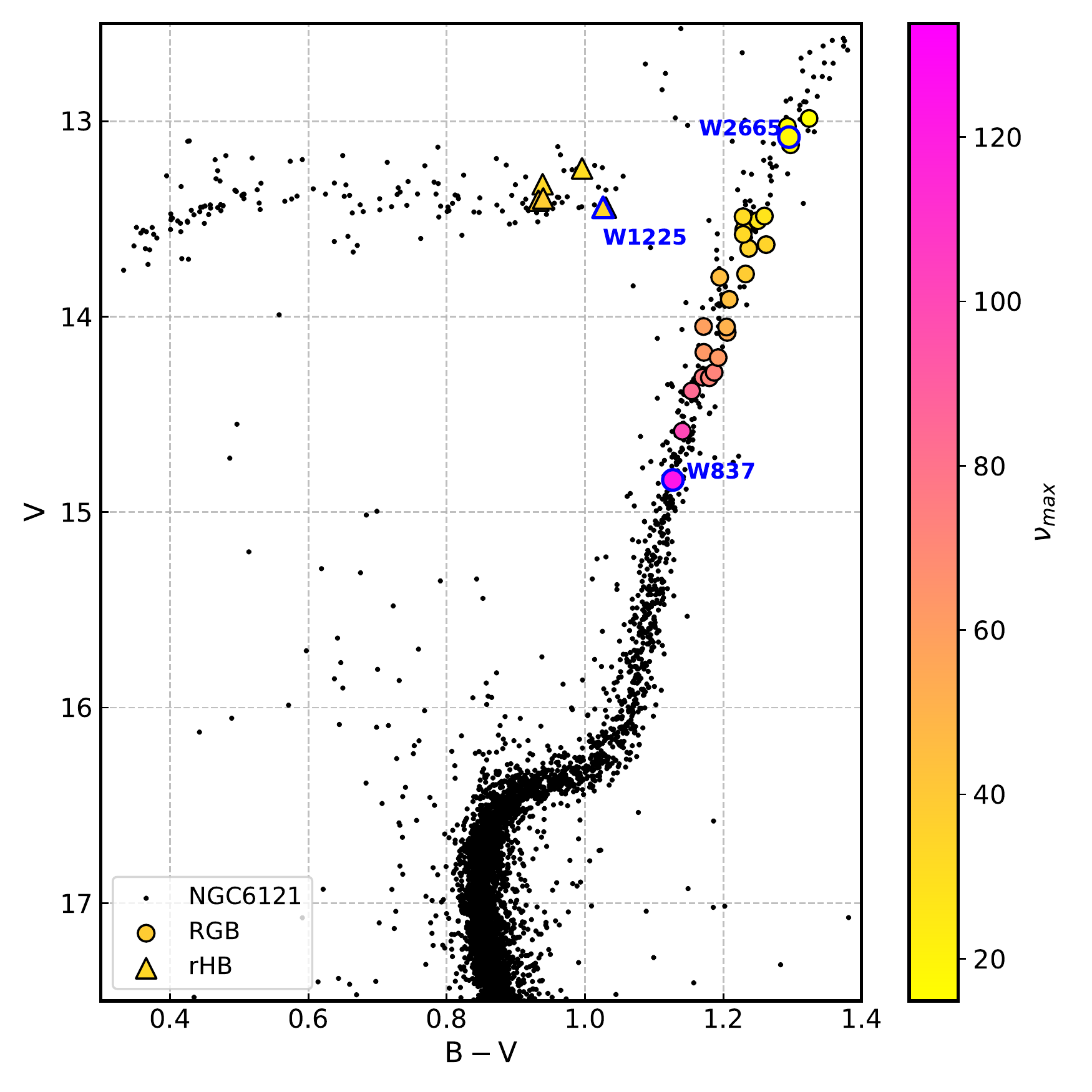}
    \caption{\textit{Left: }Background fits (red curve) overlaid on the stellar PSD (gray) for the stars W837 (lower RGB), W1225 (HB), and W2665 (upper RGB). A smoothing of the PSD using a boxcar with width set to $\Delta\nu$ is shown for reference as the black curve. The dot-dashed blue curves depict the Harvey-like components, while the horizontal dot-dashed yellow line and the dot-dashed magenta curve represent the level of the white noise and the Gaussian envelope of the oscillations, respectively. The dashed green line on top of the background fit shows the resulting fit when incorporating the oscillation envelope. We note here that the number of components in the model used for the background fit is identified as part of the procedure (see \S\,\ref{sec:obs_datasets} and \S\,\ref{sec:bkg}) \textit{Right:} Optical CMD of M4 from \citet{Stetson_2019} data in the B and V bands. We show the position of the targets with good photometry and confirmed oscillation excess. The three stars of the left panels are identified as labelled.}
    \label{fig:bkg_fit}
\end{figure*}

\begin{table*}[]
\caption{Global and asteroseismic parameters of the targets with confirmed detection. Columns are: Name of the star in \citet{Wallace19}, Star ID in the \textit{Gaia} eDR3 \citep{Gaia2021}, right ascension (RA) and declination (DEC), V magnitude from \citet{Stetson_2019} corrected for differential reddening, $\rm C_{UBI}$ index, effective temperature ($\rm T_{\rm eff}$, 100~K has been assumed as $1\,\sigma$ value). Luminosity in solar units, $\rm \nu_{max}$ and $\Delta\nu$ in $\rm \mu Hz$ and, finally, population ID either from spectroscopy ($\rm POP_{Spec}$) or using the  $\rm C_{UBI}$ index ($\rm POP_{CUBI}$). We report at the bottom the asteroseismic parameters of the targets with more uncertain photometry in the \cite{Stetson_2019} data.}
\begin{adjustbox}{width=1.99\columnwidth}
\centering
\renewcommand{\arraystretch}{1.2} 
\begin{tabular}{lccccccccccc}
\hline
\hline
  WID &     Gaia$_{\rm eDR3}$ ID &         RA (J2000) &        DEC (J2000) & $\rm V_{dr}\,(mag)$ & $\rm C_{UBI}\,(mag)$ & $\rm T_{eff}\,(K)$ &  $\rm L/L_\odot$ & $\rm \nu_{max}\, (\mu Hz)$ & $\rm \Delta \nu\, (\mu Hz)$& POP$_{\rm Spec}$& POP$_{\rm CUBI}$ \\
\hline
 W491 & 6045479078332149632 & 245.798065 & -26.444676 &              13.757 &               -2.059 &             4788 & 37.865$\pm$4.083 &           38.724$\pm$0.965 &             4.616$\pm$0.180  &2G  & 2G \\
 W508 & 6045478047540140544 & 245.800242 & -26.495482 &              14.307 &               -2.096 &             4926 & 21.765$\pm$2.192 &           75.929$\pm$2.367 &             7.613$\pm$0.231  &--  & 2G \\
 W760* & 6045477841381440384 & 245.820424 & -26.496654 &              13.129 &               -2.038 &            4668 & 71.180$\pm$7.748 &           18.148$\pm$1.024 &             2.558$\pm$0.081 &2G  &  2G \\
 W779 & 6045479490649030272 & 245.822060 & -26.416080 &              14.059 &               -2.114 &             4844 & 28.142$\pm$3.030 &           53.392$\pm$2.258 &             6.056$\pm$0.171  &--  &  1G \\
 W799 & 6045478288058130304 & 245.823516 & -26.452857 &              13.313 &               -1.915 &             5596 & 45.743$\pm$4.332 &           33.108$\pm$1.851 &             4.543$\pm$0.117  &--  & rHB \\
 W837 & 6045478081899629696 & 245.825992 & -26.485369 &              14.842 &               -2.071 &             5030 & 12.892$\pm$1.235 &          123.780$\pm$3.537 &            11.765$\pm$0.309  &--  &  2G \\
W1068 & 6045478184978847616 & 245.839217 & -26.475993 &              14.307 &               -2.076 &             4900 & 21.934$\pm$2.229 &           71.632$\pm$2.839 &             7.449$\pm$0.301  &--  &  2G \\
W1091 & 6045478666015246592 & 245.840398 & -26.446373 &              13.488 &               -2.054 &             4787 & 48.529$\pm$5.227 &           31.402$\pm$1.722 &             3.899$\pm$0.296  &NC  &  2G \\
W1156 & 6045477910100876544 & 245.843863 & -26.495077 &              14.053 &               -2.134 &             4858 & 28.109$\pm$3.026 &           50.740$\pm$0.995 &             5.375$\pm$0.272  &--  &  1G \\
W1225 & 6045477978820371712 & 245.847568 & -26.486060 &              13.440 &               -1.983 &             5326 & 43.459$\pm$4.203 &           34.333$\pm$1.907 &             5.279$\pm$0.364  &--  & rHB \\
W1582 & 6045477944460600960 & 245.866076 & -26.486544 &              14.293 &               -2.138 &             4885 & 22.325$\pm$2.271 &           72.196$\pm$2.312 &             7.132$\pm$0.440  &--  &  1G \\
W1608 & 6045478734734712832 & 245.867277 & -26.435713 &              13.228 &               -1.985 &             5421 & 51.551$\pm$4.936 &           30.400$\pm$2.277 &             4.385$\pm$0.111  &--  & rHB \\
W1717 & 6045478730423545984 & 245.872427 & -26.441030 &              14.364 &               -2.061 &             4966 & 20.374$\pm$2.016 &           83.731$\pm$1.718 &             8.619$\pm$0.680  &--  &  2G \\
W1763 & 6045503057118542208 & 245.874265 & -26.390373 &              12.973 &               -2.042 &             4616 & 84.140$\pm$9.199 &           14.920$\pm$1.461 &             2.212$\pm$0.078  &2G  &  2G \\
W1912 & 6045478764783374208 & 245.880406 & -26.421051 &              14.577 &               -2.075 &             4994 & 16.636$\pm$1.628 &          100.116$\pm$2.931 &            10.021$\pm$1.264  &--  &  2G \\
W2021 & 6045466571386686208 & 245.884849 & -26.489523 &              13.918 &               -2.140 &             4841 & 32.030$\pm$3.448 &           44.785$\pm$1.422 &             5.374$\pm$0.229  &1G  &   1G \\
W2022* & 6045478528576252288 & 245.884878 & -26.439056 &              13.024 &               -2.115 &            4677 & 77.933$\pm$8.468 &           15.986$\pm$1.047 &             2.418$\pm$0.085 &1G**&  1G \\
W2034 & 6045502305515873024 & 245.885444 & -26.409477 &              13.498 &               -2.135 &             4764 & 48.533$\pm$5.240 &           26.413$\pm$1.714 &             3.480$\pm$0.204  &--  &  1G \\
W2162 & 6045478356777481984 & 245.890859 & -26.463824 &              13.649 &               -2.096 &             4782 & 41.995$\pm$4.528 &           35.280$\pm$1.784 &             4.245$\pm$0.359  &2G  &  2G \\
W2386 & 6045478558624847488 & 245.899759 & -26.439063 &              13.440 &               -2.054 &             5314 & 43.543$\pm$4.214 &           36.400$\pm$0.726 &             5.207$\pm$0.178  &--  & rHB \\
W2665* & 6045501961918453504 & 245.911909 & -26.428551 &              13.084 &               -2.036 &            4675 & 73.824$\pm$8.007 &           17.806$\pm$0.445 &             2.526$\pm$0.089 &2G  &  2G \\
W2678 & 6045502477314585728 & 245.912264 & -26.369307 &              13.555 &               -2.128 &             4805 & 45.290$\pm$4.884 &           31.495$\pm$0.878 &             3.862$\pm$0.300  &1G  &  1G \\
W2887 & 6045466433947664256 & 245.922274 & -26.485789 &              13.418 &               -1.926 &             5614 & 41.411$\pm$3.922 &           39.876$\pm$2.520 &             3.938$\pm$0.204  &--  & rHB \\
W3033* & 6045466674465880064 & 245.929551 & -26.468741 &              13.582 &               -2.058 &            4785 & 44.566$\pm$4.798 &           32.892$\pm$1.396 &             4.102$\pm$0.334 &2G  &  2G \\
W3041 & 6045490210887898368 & 245.930070 & -26.444946 &              14.191 &               -2.168 &             4927 & 24.184$\pm$2.472 &           62.427$\pm$2.156 &             6.847$\pm$0.872  &--  &  1G \\
W3073 & 6045501996278180992 & 245.931331 & -26.427090 &              14.217 &               -2.082 &             4868 & 24.146$\pm$2.600 &           61.963$\pm$1.000 &             6.873$\pm$0.374  &--  &  2G \\
W3480 & 6045466399577821440 & 245.954138 & -26.488691 &              13.816 &               -2.096 &             4875 & 34.725$\pm$3.550 &           45.087$\pm$1.775 &             5.228$\pm$0.182  &NC  &  2G \\
W3528 & 6045466399587848704 & 245.957287 & -26.481054 &              14.070 &               -2.152 &             4929 & 27.020$\pm$2.737 &           59.026$\pm$3.100 &             6.341$\pm$0.195  &--  &  1G \\
W3564 & 6045465643673557760 & 245.959422 & -26.490526 &              13.505 &               -2.008 &             4803 & 47.534$\pm$5.117 &           31.958$\pm$2.938 &             4.029$\pm$0.230  &2G  &  2G \\
W3742 & 6045489901650196864 & 245.970856 & -26.468500 &              13.653 &               -2.157 &             4720 & 42.973$\pm$4.649 &           35.071$\pm$1.680 &             4.126$\pm$0.280  &1G  &  1G \\
W3929* & 6045490103497000320 & 245.985488 & -26.424575 &              13.428 &               -1.940 &            5595 & 41.541$\pm$3.942 &           38.242$\pm$1.957 &             4.811$\pm$0.297 &--  & rHB \\
W4488 & 6045489351894331776 & 246.061253 & -26.454663 &              13.509 &               -2.065 &             4708 & 49.304$\pm$5.316 &           27.586$\pm$1.425 &             4.324$\pm$0.146  &--  &  2G \\
\hline
\multicolumn{11}{c}{Targets with more uncertain photometry}\\
\hline
W1512 & 6045479593728227072 & 245.863124 & -26.40367 &             -- &               -- &             -- & -- &           39.516$\pm$1.704&             5.401$\pm$0.154 &--  &  NO \\
W3079 & 6045490210887906176 & 245.931592 & -26.43842 &             -- &               -- &             -- & -- &           34.562$\pm$1.228&             3.852$\pm$0.173 &--  &  NO \\
W3474 & 6045466399587823744 & 245.953973 & -26.48835 &             -- &               -- &             -- & -- &           44.461$\pm$2.358&             5.151$\pm$0.292 &--  &  NO \\
W4092 & 6045490485765778560 & 245.998091 & -26.42664 &             -- &               -- &             -- & -- &           28.065$\pm$2.369&             3.729$\pm$0.157 &--  &  NO \\
W4283 & 6045490447094400384 & 246.024697 & -26.42321 &             -- &               -- &             -- & -- &           50.277$\pm$1.645&             5.849$\pm$0.232 &--  &  NO \\
\hline
\hline
\multicolumn{12}{l}{1G = first generation,  2G = second generation, NC = Not Certain, rHB = Red Horizontal Branch Star}\\
\multicolumn{12}{l}{* also in \citet{Miglio_2016}, ** Diverging identification: 1G in \citet{marino_2008}, 2G in \citet{carretta_2009}, we agree with the former}
\end{tabular}
\end{adjustbox}
\label{tab1}
\end{table*}

We computed a Power Spectral Density (PSD) for each star for which light curves from \cite{Wallace19} were available. This procedure was carried out for a total of 4554 stars by means of the \textsc{kadacs} libraries \citep[e.g.][]{Garcia11}. We performed a first skimming of the potential candidates by visually inspecting the PSDs and checking their location in the colour-magnitude diagram (CMD). Our final sample includes 54 stellar candidates located on the RGB and in the red part of the HB (rHB). We then fitted each PSD by means of the background modeling technique presented by \cite{Corsaro14,Corsaro15cat,Corsaro17metallicity} based on the \textsc{Diamonds} code. The background model adopted for each star was selected through a Bayesian model comparison process. 

In addition to fitting the background, we obtain the position of the Gaussian envelope of the stellar oscillations and thus the frequency of maximum power, $\nu_\mathrm{max}$, for each star, while the large frequency separation, $\Delta\nu$, was estimated from the auto-correlation function (ACF) over the range of the PSD that contains the stellar oscillations. We have confirmed detection in 37 stars, 6 of which located in the rHB.
Three examples of the PSD fit are shown in the left part of Fig.\,\ref{fig:bkg_fit}. More details on the procedure are reported in \S\,\ref{sec:bkg}. The determined global asteroseismic parameters ($\nu_\mathrm{max}$ and $\Delta\nu$) are listed in Table \ref{tab1}.

We used stellar astrometry from \textit{Gaia} and photometry in the $U$, $B$, $V$ and $I$ bands from \cite{Stetson_2019} to better characterize our targets. We identified cluster members by using the method by \cite{Cordoni_2018}, which is based on \textit{Gaia} eDR3 parallaxes and proper motions \citep{Gaia2021}.  Photometry has been corrected for differential reddening as in \citet{milone_2012c}. In the right panel of Fig.~\ref{fig:bkg_fit}, we plot the colour-magnitude diagram in $B-V$ vs $V$ bands in which we highlight the selected targets,each of them coloured according to its $\nu_\mathrm{max}$ value. Our sample significantly extends in number and $\nu_\mathrm{max}$ range the sample in \citet{Miglio_2016}. Five stars in our sample have uncertain photometry\footnote{Point-like sources that are well-matched by the Point-spread-function model follow a well-defined trend in the magnitude versus sharpness \citep{Stetson_2019} plane. We excluded from the analysis all stars with large sharp values with respect to the bulk of stars with similar magnitudes.}, and therefore we remove them from our analysis. 

We exploit the capabilities of the $C_{U,B,I}=(U-B)-(B-I)$ index \citep{Milone_2013,Monelli_2013} to separate the two stellar populations in this cluster. In Table~\ref{tab1}, we report the values of corrected $V$ magnitude and the value of the $C_{U,B,I}$ index for all our targets with good photometry. The correct identification of the generation which our RGB stars belong is necessary to correctly assess the integrated mass loss for the stars in this GC because our rHB stars all belong to the first generation \citep[see][and the first part of \S \ref{sec:multi_pop_mass}]{marino_2011}.
After the selection procedure, our final sample contains 26 RGB stars (16 second generation or 2G,  10 first generation or 1G) and 6 rHB stars.

A set of temperatures was derived from $V-I$ colours as these are less affected by the shift in chemistry between the stellar populations in M4. In order to explore possible systematic uncertainties introduced by the choice of bands, we derive a second set of temperatures from the $B-V$ colour. We adopted $E(B-V)=0.37$ and $E(V-I)=0.53$ from \citet{Hendricks2012}. To obtain the values of $T_\mathrm{eff}$ for both combinations of bands, we used the bolometric corrections and the colour--$T_\mathrm{eff}$ relation from \citet{Casagrande2014}, adopting [Fe/H]=-1.1 and [$\alpha$/Fe]=+0.4 \cite[][]{marino_2008}.  An iterative procedure was adopted to refine the results by using the asteroseismic surface gravity obtained from $\nu_\mathrm{max}$ and $T_\mathrm{eff}$.
For consistency with the adopted colour-$T_\mathrm{eff}$ relations, we use the distance modulus $(m-M)_0=11.20\pm0.1$ from \citet{Casagrande2014}\footnote{This is compatible with the one derived in \cite{Hendricks2012} within 1~$\sigma$.}, which combined with the magnitudes of our stars provide us with their luminosity, and hence, by using the derived effective temperature, their photospheric radius ($R_\mathrm{CMD}$).

\section{Masses and radii}
\label{sec:mass_radius}

Combining the above decided asteroseismic and non-asteroseismic parameters, we estimate stellar masses and radii with classical scaling relations. Combining $\Delta\nu\propto \rho^{1/2}$ and $\nu_{max}\propto gT_{\rm eff}^{-1/2}$ \citep[see e.g.][]{BrownTM91,Frandsen2007,Chaplin2013,Miglio_2012,Miglio_2016} with $L\propto R^2T_{\rm eff}^4$ we obtain the set of equations reported below:
\begin{equation}
  \rm   \left(\frac{M_1}{M_\odot}\right) \simeq 
  \left( \frac{\nu_{max}}{\nu_{max,\odot}}\right)^{3} 
  \left(\frac{\Delta \nu}{\Delta \nu_{\odot}}\right)^{-4} 
  \left( \frac{T_{eff}}{T_{eff,\odot}}\right)^{3/2},
  \label{eq:scal1}
\end{equation}

\begin{equation}
  \rm   \left(\frac{M_2}{M_\odot}\right) \simeq 
  \left(\frac{\Delta \nu}{\Delta \nu_{\odot}}\right)^{2} 
  \left( \frac{L}{L_{\odot}}\right)^{3/2}
  \left( \frac{T_{eff}}{T_{eff,\odot}}\right)^{-6},
  \label{eq:scal2}
\end{equation}

\begin{equation}
  \rm   \left(\frac{M_3}{M_\odot}\right) \simeq 
  \left( \frac{\nu_{max}}{\nu_{max,\odot}}\right) 
  \left( \frac{L}{L_{\odot}}\right)
  \left( \frac{T_{eff}}{T_{eff,\odot}}\right)^{-7/2},
  \label{eq:scal3}
\end{equation}

\begin{equation}
  \rm   \left(\frac{M_4}{M_\odot}\right) \simeq 
  \left( \frac{\nu_{max}}{\nu_{max,\odot}}\right)^{12/5}
  \left(\frac{\Delta \nu}{\Delta \nu_{\odot}}\right)^{-14/5} 
  \left( \frac{L}{L_{\odot}}\right)^{3/10}.
  \label{eq:scal4}
\end{equation}

$\Delta\nu$ is not taken at face value: multiplicative corrections to the relation between $\Delta\nu$ and $\rho^{1/2}$ ($f_{\Delta\nu}$) are applied following the approach described in \citet{Rodrigues_2017} which, so far, has yielded masses and radii showing no systematic deviations to within few percent of independent estimates (see e.g. \citealt{Miglio_2016}, \citealt{Rodrigues_2017}, \citealt{Handberg_2017},  \citealt{Brogaard_2018}, who partially revisited the work by \citealt{Gaulme_2016}, and \citealt{Buldgen_2019}). Our computations of $f_{\Delta\nu}$ are described in  \S\,\ref{sec:sys_correction} while the full set of parameters is reported in Table \ref{tab2}.

We may also estimate the stellar radii with:
\begin{equation}
  \left(\frac{R_\mathrm{S}}{R_\odot}\right) \simeq 
  \left( \frac{\nu_\mathrm{max}}{\nu_\mathrm{max,\odot}}\right)
  \left(\frac{\Delta \nu}{\Delta \nu_{\odot}}\right)^{-2} 
  \left( \frac{T_\mathrm{eff}}{T_\mathrm{eff,\odot}}\right)^{1/2}.
    \label{eq:scal5}
\end{equation}
For solar reference values, we adopt $\rm \Delta\nu_\odot=135.1\,\mu Hz$, $\rm \nu_\mathrm{max,\odot}= 3090\,\mu Hz$, and $T_\mathrm{eff,\odot}=5777\,K$ \citep{Huber2013}. Finally the $1\sigma$ interval on each quantity has been obtained with standard error propagation methods, starting from the $1\sigma$ uncertainty of the parameters involved in each equation. 

As an independent check for potential biases in the seismically inferred masses and radii, we compare the asteroseismic radius ($R_\mathrm{S}$) and the photometric one of all of our targets. For RGB stars, we find that  $\langle(R_\mathrm{S}-R_\mathrm{CMD})/\sigma\rangle=0.11\pm0.15$ of the combined error. This lends confidence to both the mass and distance scale, given the strong correlation between seismic masses and radii  \citep[see e.g.][ and references therein]{basu2011,wu2014,Khan_2019,Miglio2021}. For  rHB stars, however, the situation is  different as we find significant differences in radii. For these stars we have $\langle(R_\mathrm{S}-R_\mathrm{CMD})/\sigma\rangle=1.58\pm0.77$ and at the same time the mass of our HB star calculated with the scalings involving $\Delta\nu$ (Eq. \ref{eq:scal1},\ref{eq:scal2},\ref{eq:scal4}) do not agree with each other.
A thorough exploration of possible systematic (see \S\,\ref{sec:syst_all}) points to an issue related to measuring $\Delta\nu$ in rHB.

We checked the robustness of our $\Delta\nu$ measurement via a simple test with the \textit{Kepler} data. We degraded the entire 4yr duration of the \textit{Kepler} observations for 2 rHB stars -- namely KIC~8694070 and KIC~11450315 -- by diving the full time series into 80-day chunks and we derived $\Delta\nu$ from these shorter data sets. We find that the measured $\Delta\nu$ fluctuates up to 40\%, leading to overestimate (or underestimate) the mass and the radius of the HB stars up $\sim8$ and $\sim3$ times, respectively. We give more details on this in \S\,\ref{sec:dnu}. We believe that this is due to the intrinsic complexity of the HB oscillation spectra (Matteuzzi et al in prep.) which combined with the short duration of K2 observations make accurate measurements of $\Delta\nu$ difficult. This kind of instability is not present in RGB stars that on the other hand present no issues when measuring $\Delta\nu$. Therefore, when discussing rHB stars we will use Eq. 3, which does not include $\Delta\nu$ and this unstable behaviour, to estimate the stellar parameters of rHB stars. A thorough investigation of systematic effects (in $T_\mathrm{eff}$, distance, $\Delta\nu$ and $\nu_\mathrm{max}$) is presented in \S\,\ref{sec:syst_all}.

\subsection{Mass of the RGB stars in M4} 
\label{sec:res_mass}

\begin{figure}
    \centering
    \includegraphics[width=\columnwidth]{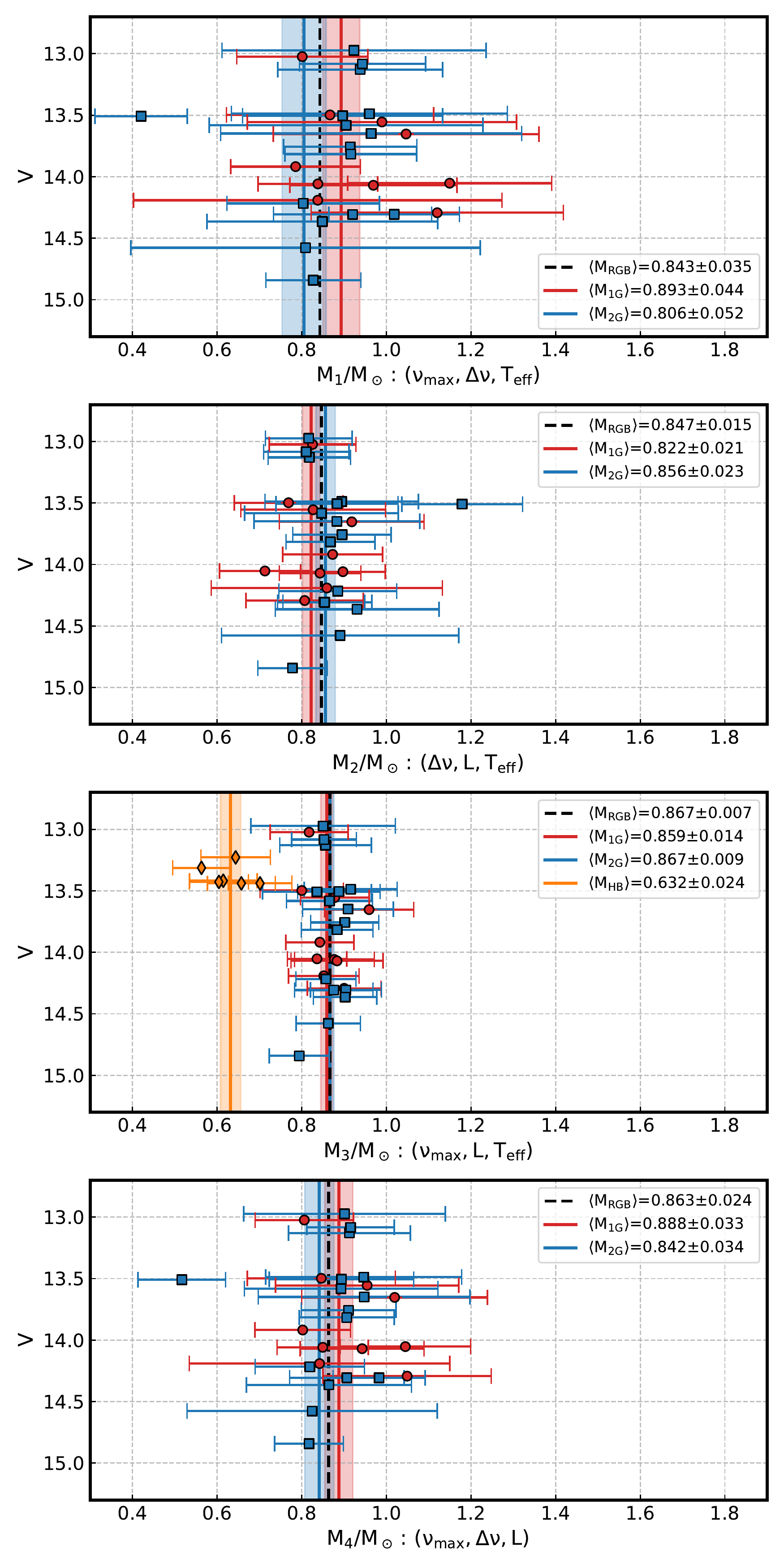}
    \caption{Mass of our targets estimated with Eq. \ref{eq:scal1} to \ref{eq:scal4}, in descending order. The black, dashed lines represent the average mass of the RGB population, while the coloured ones refer to the same quantity for the corresponding groups.}  
    \label{fig:masses}
\end{figure}

We report the mass of our RGB targets in the four panels of Fig.\,\ref{fig:masses} as the collection of red and blue points, for 1G and 2G stars respectively (see \S\,\ref{sec:multi_pop_mass} for more details). The weighted average mass for our 26 RGB stars ($\langle M_1\rangle = 0.842\pm0.035 M_\odot$, $ \langle M_2\rangle = 0.847\pm0.015 M_\odot$, $\langle M_3\rangle = 0.867\pm0.007 M_\odot$ and $\langle M_4\rangle = 0.863\pm0.024 M_\odot$) are shown as the black dashed lines in each panel. Combining these values with the isochrones from \cite{tailo_2016a,tailo_2020} we found an age in the range 11 -- 12 Gyr, with the exact value depending on the adopted equation for the RGB mass\footnote{Alternative estimates can be obtained using Basti \citep{Hidalgo2018,Pietrinferni2021} and Padova isochrones \citep{Bressan2012}, obtaining age values in similar ranges.}. These four values of mass and age agree with other published values for this cluster \citep[][]{dotter_2010,Kaluzny2013,VandenBerg_2013,Miglio_2016,tailo_2019a,jang_2019}.
Note that the different expressions linking the seismic parameters with the classical ones involve the value of $\Delta\nu$, $\nu_\mathrm{max}$, $T_\mathrm{eff}$ and luminosity ($L$) with different powers. This is the main reason for the difference in the precision with which the stellar parameters are derived with the different scaling relations. 
Within the current uncertainties, no mass variation is observed over the magnitude range examined. 

We also estimate the mean mass of 1G and 2G RGB stars, however, we do not find a significant mass difference given the current systematic and statistical uncertainties (see \S\,\ref{sec:multi_pop_mass} and \ref{sec:syst_all} for more details). This is not surprising as the predicted theoretical mass difference between the two populations is estimated to be $0.017\ \rm M_\odot$ \citep{tailo_2019a}, with the 1G being composed of more massive stars. This mass difference has been obtained assuming that the two populations are strictly coeval, therefore it originates only from the effects chemical enrichment has on evolution timescale. However, all the most successful scenarios for multiple stellar populations formation predict that the two generations of stars form within a small interval of time, compared to the age of the cluster. Such interval should be few $\rm 10^8 yr$, at the largest, depending on the scenario \citep[see][]{renzini_2015,dantona_2016,bastian_2018,kim_2018}.  Therefore, when one takes into consideration that the 2G is the youngest, the mass difference reduces. Our results go in this direction; however our uncertainties prevent us to say more.

\subsection{The mass of the HB stars}
\label{sec:hb}

\begin{figure}
    \centering
    \includegraphics[width=0.65\columnwidth]{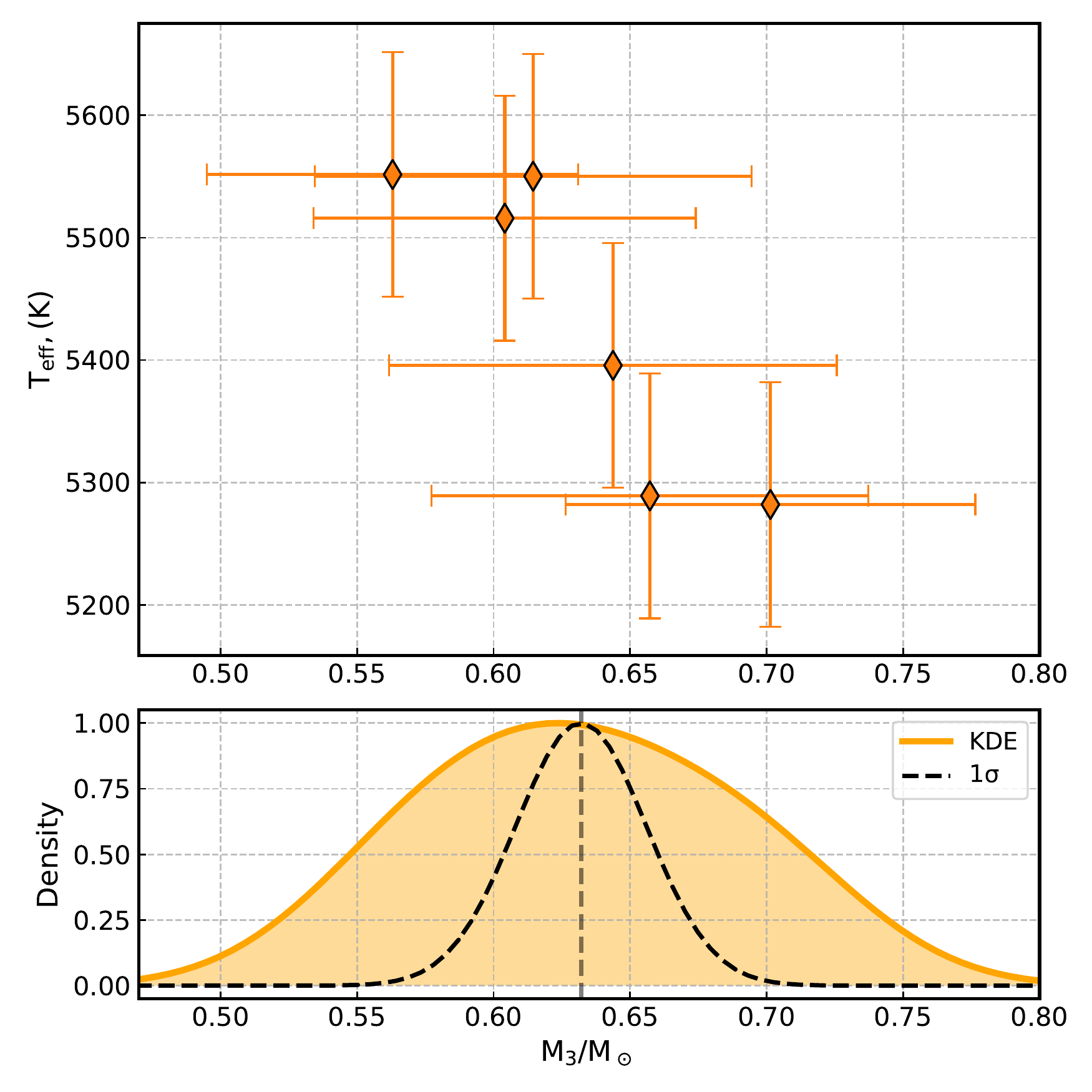}
    \caption{\textit{Top:} Temperature of our rHB targets as function of their mass. \textit{Bottom:} Gaussian Kernel distribution for the data (orange). The black dashed line represent a single-mass population centered a $\rm\langle M_{HB} \rangle$ with its corresponding sigma. }
    \label{fig:HB}
\end{figure}

The masses for our 6 HB stars are reported as the orange diamonds in the third panel of Fig.\ref{fig:masses}. The average value is  $\langle M_{HB} \rangle=0.632\pm0.024\,M_\odot$, in good agreement with other estimates of the average rHB mass for this cluster \citep[][]{tailo_2019a,jang_2019}. Even if the number of targets is small, the scatter shown by these values does not seem compatible with a single mass. This is shown in lower panel of Fig.~\ref{fig:HB}, where we compare the Gaussian kernel-density distribution of our stars (orange) with a single mass distribution (dashed black), centered at the $\langle M_{HB} \rangle$ with the corresponding $\sigma$.

An inspection of the properties of these stars reveals the presence of a clear trend between mass and effective temperature (top panel of Fig.\,\ref{fig:HB}), with an estimated Spearman $r$ index, $r\sim -0.94\, (p<0.005)$, indicating that this gradient is significant at $\sim 3\sigma$ level. We note that this behaviour is consistent with theoretical models, which predict hotter stars for smaller masses \citep[with a smaller H envelope mass, see][ for instance]{Girardi2016}, and, in such a case, this would be the first direct observation of HB extension in terms of mass variations.

\subsection{Integrated mass loss}
\label{sec:mloss}

\begin{figure*}
    \centering
    \includegraphics[width=0.85\columnwidth]{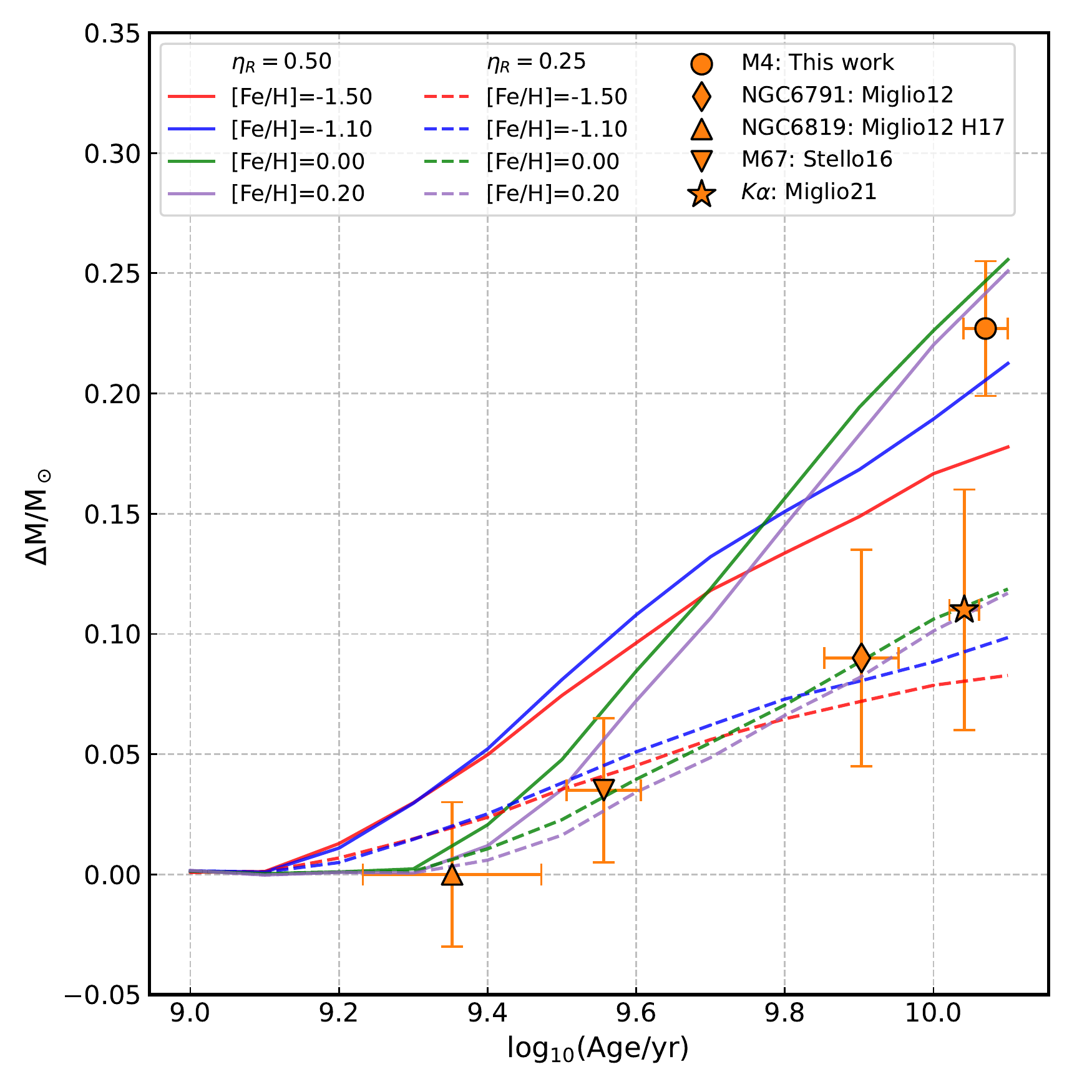}
    \includegraphics[width=0.85\columnwidth]{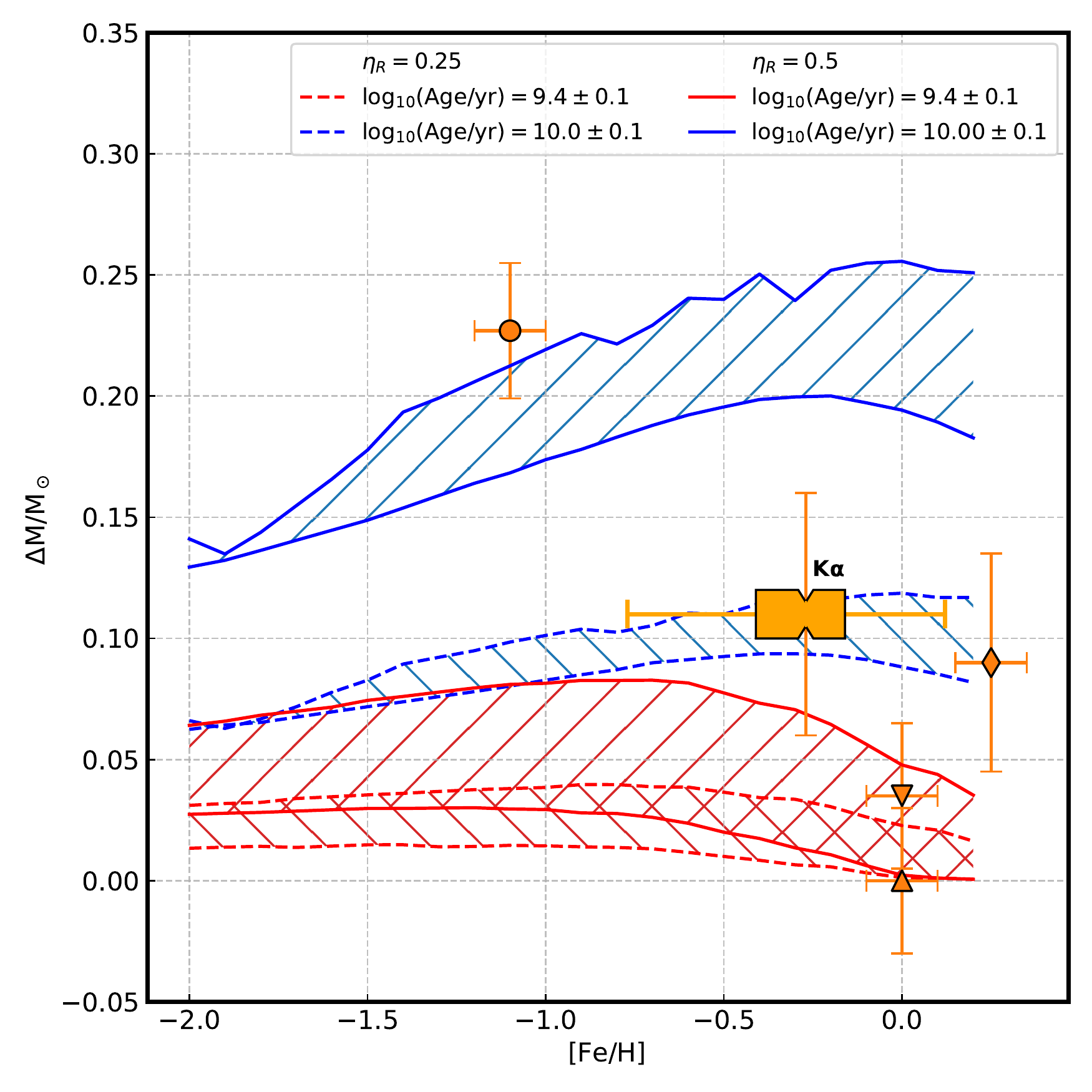}
    \caption{Mass loss measured using asteroseismic observables for M4, NGC6791 \citep[][]{Miglio_2012}, NGC6819 \citep[Miglio et al. 2012 and][H17]{Handberg_2017}, M67 \citep[][]{Stello2016} and the $\alpha$-enhanced stars in the \textit{Kepler} field \citep[$K\alpha$,][]{Miglio2021} as funcion of age (left) and  metallicity (right). Two sets of models for different \citep{reimers_1975} mass loss parameter, $\rm \eta_R=0.25,0.50$, are included as comparison.}
    \label{fig:mloss}
\end{figure*}

The average integrated RGB mass loss can be obtained by subtracting the average mass of the rHB stars from that of the RGB ones.  To get the correct value of integrated mass loss, the rHB stars have to be compared with their progenitor in the 1G, along the RGB (which have an average mass of 0.859 $M_\odot$, see \S\,\ref{sec:multi_pop_mass}). The estimated average integrated mass loss of the 1G stars is therefore $\Delta M=0.227\pm0.028 M_\odot$, compatible with the one in \citet[][$\Delta M=0.209\pm0.024 M_\odot$]{tailo_2019a}. If we convert this value into  a mass loss rate \citep[as in Fig 16 in][]{tailo_2020}, we obtain that the \cite{reimers_1975} mass loss parameter associated is $\rm \eta_{R}\sim0.48$, while using Padova stellar models \citep{Bressan2012} we obtain a value closer to 0.50.

Adding this result to the previous determinations of mass loss rate in stellar clusters using asteroseismology \citep{Miglio_2012,Stello2016,Handberg_2017,Miglio2021} allows us to cover a wide range in age and metallicity.  In Fig.\,\ref{fig:mloss}, we show the mass loss expected from stellar models from the Padova database \citep{Bressan2012}, with different chemical composition, age, and two values of the \citet{reimers_1975} mass loss parameter $\rm \eta_{R}= 0.25; 0.50$. These predictions are compared with the integrated mass loss estimated from asteroseismic studies of stellar clusters (including the value for M4 derived in this paper) and of the high-[$\alpha$/Fe] red giants in the \textit{Kepler} field \citep{Miglio2021}. 
The agreement between the asteroseismic estimates of mass loss and the estimate coming from fitting the photometric HB data with stellar populations models \citep[]{tailo_2019a,jang_2019} corroborates the hypothesis from \cite{tailo_2020,Tailo2021} that mass loss in old GC is a substantially different phenomenon than in open clusters and in field stars which, if one assumes Reimers’ parameterisation, is properly described only invoking a higher mass loss parameter. 

This divergent behaviour could be connected to different environmental factors (interaction with the Galaxy or different formation environment), be the footprint of early dynamical interaction between the stars in GCs (stellar collisions and binaries), be the product of some still poorly understood physics inside RGB stars \citep[like the mechanism proposed in][]{Fusi-Pecci1975,Fusi-Pecci1976}, or even a combination of all the previous. Indeed, the extensive study performed in Tailo et al. (2020,2021) shows that the large population of GCs analysed lies on a relation systematically higher than the one for the open clusters and the field stars of comparable metallicity (where the comparison is possible). Furthermore, said relation is compatible with the ones obtained for dwarf spheroidal galaxies \citep[e.g.][]{salaris_2013,savino_2019}. This suggest that RGB mass loss in old stellar association is somewhat universal.

\section{Conclusions}

This study presents a new asteroseismic analysis of targets located in M4 that features the largest sample in this kind of studies for GCs (37 stars). We combined asteoseismic measurements of $\nu_\mathrm{max}$ and $\Delta \nu$ with the \cite{Stetson_2019} photometric catalogue to provide a direct measurement of 32 stars, 6 of which are rHB stars. 

Our results confirm the validity of asteroseismology in the low-metallicity regime, thus giving more solidity to the results in \citet{Valentini2019,montalban_2021,Miglio_2016,Miglio2021}, and providing a direct estimate of the RGB mass loss in this cluster which is in agreement with the estimate derived from the fitting of the HB photometric data with stellar populations models \citep{tailo_2019a,jang_2019}. Although we separated the stellar populations in this cluster efficiently, our uncertainties prevent us from telling anything about the mass difference between the two stellar populations. However, this work paves the way for future studies using a more detailed asteroseismic analysis of the individual mode frequencies that will allow for more precise mass estimates \citep[see e.g.][]{montalban_2021}, less reliant on the method used to obtain $T_\mathrm{eff}$. Finally, the gradient we observe in our HB stars is the first, direct and model-independent observation showing the colour-temperature-mass correlation known to exists in HB stars from the theoretical models.

\begin{acknowledgements}
MT, AM, JM, AS, GC acknowledge support from the ERC Consolidator Grant funding scheme (project ASTEROCHRONOMETRY, https://www.asterochronometry.eu, G.A. n. 772293). APM acknowledges support from the European Union’s Horizon 2020 research innovation programme (Grant Agreement ERC-StG 2016, No 716082 ’GALFOR’, PI: Milone). Funding for the Stellar Astrophysics Centre is provided by The Danish National Research Foundation (Grant agreement No.~DNRF106). E.C. acknowledges support from PLATO ASI-INAF agreement no. 2015-019-R.1-2018.
\end{acknowledgements}

%
  \bibliographystyle{aa} 
  \bibliography{paper_m4_astero} 
%



\begin{appendix} 

\section{Obtaining the global asteroseismic parameters}
\label{sec:bkg}

For assessing the presence of an oscillation power excess, we computed a Power Spectral Density (PSD) for each star for which light curves from \cite{Wallace19} were available. This procedure was carried out for a total of 4554 stars by means of the \textsc{kadacs} libraries \citep[e.g.][]{Garcia11}. A skimming of potential candidates was done by a visual inspection of the PSDs in conjunction to the location of the star in the CMD. We ended up with a list of 54 stellar candidates, with evolutionary stages spanning from the low part of the red giant branch (RGB) to the horizontal branch (HB).

We proceeded by performing an actual fit to each PSD of the candidate stars by means of the background modeling technique presented by \cite{Corsaro14,Corsaro15cat,Corsaro17metallicity}, which involves the use of the public code \textsc{Diamonds+Background} based on the Bayesian nested sampling Monte Carlo algorithm \citep{Skilling04}. The background model adopted for each star was selected through a Bayesian model comparison process that uses the Bayesian evidence as a statistical quantity for model selection. The Bayesian evidence is a direct output of a statistical inference performed by means of the \textsc{Diamonds} code. As part of its outputs, the Bayesian procedure also indicates the number of Harvey-like component \cite{Harvey1985} needed to fit the background of the PSD. 
We have found that either a model consisting of a single or a double Harvey-like component represents an adequate choice for the data sets provided by \cite{Wallace19}.
In this case, these Harvey-like components mostly refer to the presence of a granulation-activity signal. On top of the Harvey-like components we have included a Gaussian function to mimic the presence of a power excess caused by stellar oscillations. The location of the power excess was initialized by means of an input guess for the frequency of maximum oscillation power, $\nu_\mathrm{max}$, which was extrapolated by combining information from a visual inspection of the PSD and the position of the star in the CMD, hence its $\log g$. An example of the fit produced is shown in Fig.~\ref{fig:bkg_fit} for stars in three different locations of the CMD. Here we can observe how the number of significant Harvey-like components decreases from two to one as the oscillation power excess moves to lower frequencies. The limited frequency resolution of the datasets increasingly penalizes our capability to infer the background properties because the usable frequency range to fit a single Harvey-like profile shrinks as the star evolves along the RGB. 

The presence of an oscillation power excess for the stars that were considered as potential candidates was further confirmed through the adoption of a Bayesian model comparison, once again performed by means of the \textsc{Diamonds} code. This procedure was carried out by comparing the Bayesian evidence of a model that incorporates the Gaussian envelope on top of the Harvey-like functions, and of another model that only contains the latter ones. We considered as the minimum for a detection in a single star that of a weak evidence condition \citep{Trotta08}, corresponding to $\ln B_\mathrm{01} \simeq 1.0$, where $\ln B_\mathrm{01}$ is the natural logarithm of the Bayes' factor, i.e. the ratio between the Bayesian evidence of the models with and without the oscillation envelope. We have detected solar-like oscillations in 37 stars, with evolutionary stages spanning from the RGB to the HB. 

\subsection{Measuring $\Delta\nu$}
\label{sec:dnu}

\begin{figure}
    \centering
    \includegraphics[width=\columnwidth]{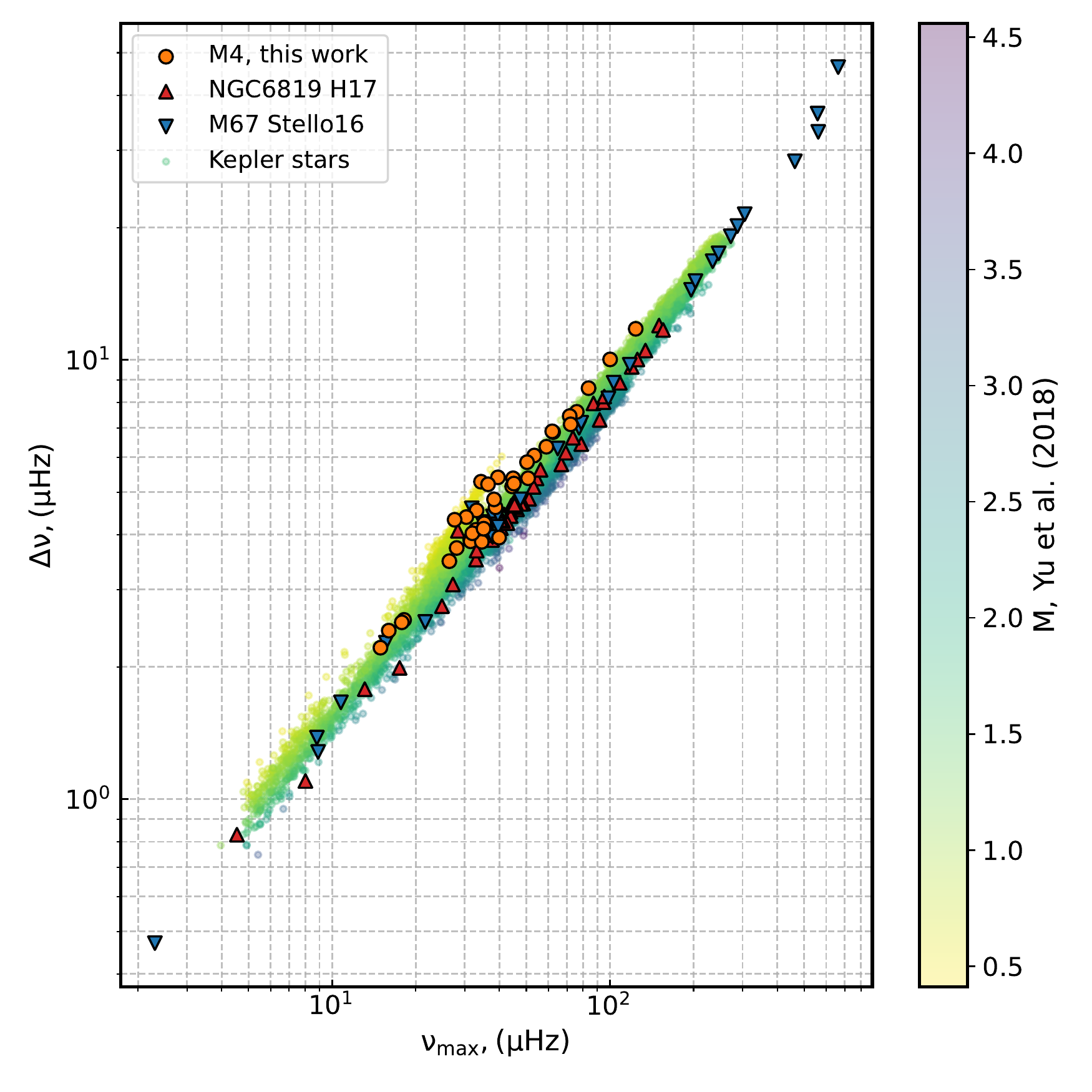}
    \caption{$\nu_{max}$---$\Delta\nu$ relation for our targets divided by groups. We compare our targets with the \textit{Kepler} stars, which are colour coded according their mass \citep[][]{Yu2018}, and the stars in NGC\,6819 from \citet[][H17]{Handberg_2017} and M67 from \citet{Stello2016}. }
    \label{fig:comp_kep}
\end{figure}

\begin{figure}
    \centering
    \includegraphics[width=\columnwidth]{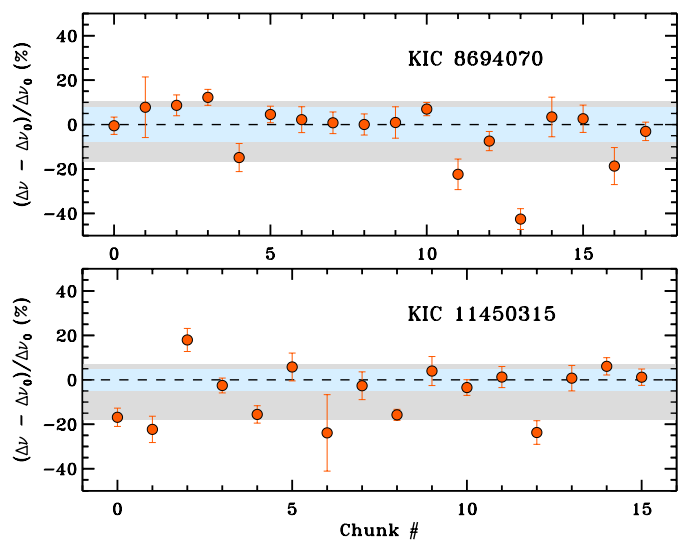}
    \caption{Deviations in percentage of $\Delta\nu$ for individual chunks each corresponding to 80 days-long time series of the core-He-burning stars KIC 8694070 (top) and KIC 11450315 (bottom) observed by \textit{Kepler}. The reference value of $\Delta\nu$, $\Delta\nu_0$, is obtained from the full-length light curve. Measurements of $\Delta\nu$ obtained from individual chunks are shown by orange dots, with associated 1-$\sigma$ error bars. The horizontal dashed line corresponds to $\Delta\nu_0$, with its 1-$\sigma$ confidence region marked by the teal shading. The gray shading indicates the 1-$\sigma$ confidence region of the sample of chunk measurements.}
    \label{fig:dnu_test}
\end{figure}

On top of testing the detection for solar-like oscillations, the fitting of the background lets us obtain the value of $\nu_\mathrm{max}$ for each star. However, the estimation of the large frequency separation $\Delta\nu$ requires that we additionally compute an auto-correlation function (ACF) over the range of the PSD that contains the stellar oscillations. This procedure is carried out by adopting a search range in $\Delta\nu$ that is centered around an input guess based on $\nu_\mathrm{max}$ through the $\Delta\nu$-$\nu_\mathrm{max}$ relation by \cite{Stello09}, and having an extent of about $\pm 50\,\%$ with respect to the input guess. In most cases, the ACF produces a clear peak that is relatively close to the input guess from the empirical relation. For completeness in Fig. \ref{fig:comp_kep} we compare the $\nu_{\rm max}$---$\Delta\nu$ relation for our targets with the one for the \textit{Kepler} stars. In the plot the latter are colour coded according to their mass from \cite{Yu2018}. We also included the stars from the open clusters M67 \citep{Stello2016} and NGC\,6819 \citep{Handberg_2017}.

However, for some of the stars belonging to the HB, the obtained $\Delta\nu$ value may deviate significantly (even up to about 40\,\%) from what one would expect. Since HB stars have more complex oscillation spectra than their RGB counterparts, which causes a less clear $\Delta\nu$ signal in the ACF, we decided to evaluate whether these deviations can be affected by systematics related to the relatively short time-span of the observations. For this purpose we conducted a test involving two red HB stars that were observed by NASA \textit{Kepler} for more than 4 years and for which a reliable estimate of $\Delta\nu$ can be obtained, namely KIC~8694070, and KIC~11450315. We created multiple realizations of their PSDs by dividing their light curves into chunks having $80\, d$ length each to simulate the observations from \cite{Wallace19}. We obtained a total of 18 chunks for KIC~8694070 and 16 chunks for KIC~11450315, as shown in Fig.~\ref{fig:dnu_test}. We then fitted the background signal in the PSD of each chunk following the same approach presented in Sect.~\ref{sec:bkg}, hence measured $\Delta\nu$ accordingly. The results plotted in Fig.~\ref{fig:dnu_test} show that the individual $\Delta\nu$ measurements suffer large scatter compared to the reference value (that obtained from the full-length light curve), although on average this behavior is not significant when the sample dispersion is taken into account. Most important, however, is that the underestimation (overestimation) can be up to about 40\,\% (20\,\%) of the reference $\Delta\nu$. This in turn implies that the corresponding asteroseismic radius may be overestimated (underestimated) up to a factor of about three, which is in qualitative and quantitative agreement with what we obtain by comparison to the CMD estimates of the stellar radius for some of our cluster HB stars. 

In addition to this issue relating to the HB stars, we find worth to mention the possibility of a second effect that intervenes when measuring $\Delta\nu$ of the RGB stars. In fact there could be a small systematic shift in the value obtained with ACF compared to the one obtained by fitting individual frequencies \citep[when the accuracy is high enough][]{Viani_2019}. According to Figure 1 in \cite{Khan_2019} this shift is $\simeq +1\%$, and therefore falling into the possible systematic effects discussed in \S\,\ref{fig:syst_all}.

\section{Multiple populations of stars in M4}
\label{sec:multi_pop_mass}

\begin{figure}
    \centering
    \includegraphics[width=\columnwidth]{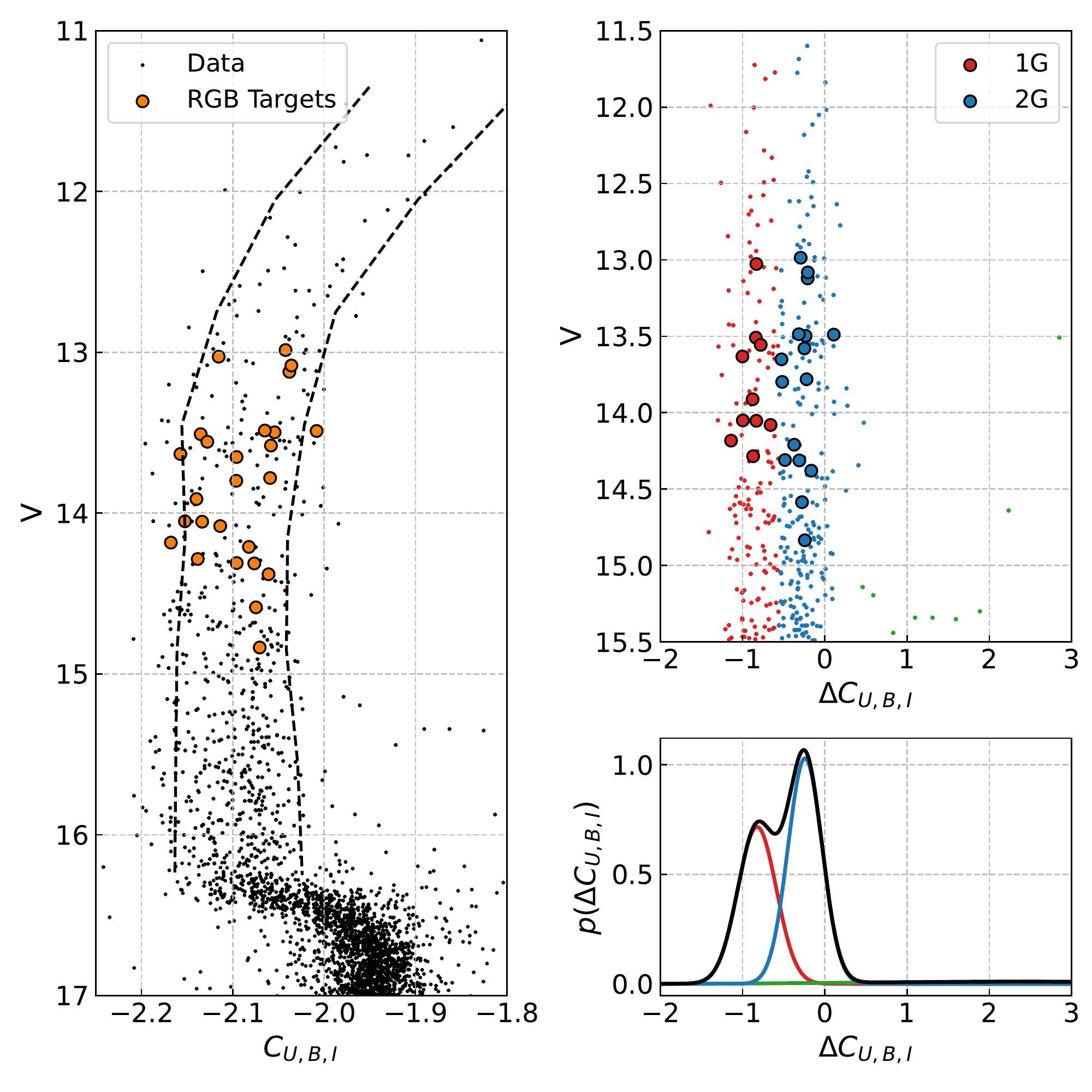}
    \caption{Summary of the method used to separate the two stellar populations along the RGB of M4. \textit{Left panel:} The photometric data in the $\rm C_{U,B,I}$ vs. $V$ plane. Our targets are highlighted as orange points. The two dashed lines mark the location of the 10th and 90th percentiles of the RGB stars, respectively for the left and the right line. \textit{Upper right panel:} The verticalized RGB ($\rm \Delta C_{U,B,I}$ vs. $V$) of M4. The points are divided between 1G (red) and 2G (blue) according the GMM reported in the bottom right panel. The green points, seen far from the main RGB, represent the binary population. \textit{Bottom right panel:} The GMM model used to identify the two stellar populations. The two most prominent Gaussian curves are the ones for the 1G and 2G stars.}
    \label{fig:populations}
\end{figure}

The main feature of GCs, such as M4, is the fact that they host a collection of different stellar populations, differing in light elements and helium abundances \citep{renzini_2015,bastian_2018,Gratton_2019}. Generally the populations in a GC divide into two groups: a first generation (1G), having light element abundances compatible with the ones of the field stars of similar metallicity (high [O/Fe] and [Mg/Fe], low [C+N/Fe], [Na/Fe], [Si/Fe] and [Al/Fe], with the standard helium abundance for its metallicity), and a second generation \citep[2G, generally divided into other sub-groups, see][and references therein]{carretta_2009,milone_2015,milone_2017,bastian_2018,milone_2018,marino_2019,Gratton_2019} showing radically different chemical patterns \citep[lower {[}O/Fe{]} and {[}Mg/Fe{]}, higher {[}C+N/Fe{]}, {[}Na/Fe{]} and {[}Al/Fe{]}, with helium enhancement that can reach 0.35 in some cases, see an example for NGC\,2808 in Fig. 2 of][and references therein]{Gratton_2019}. 
It is therefore natural to check what populations our stars belong to, and if we are able to detect a significant mass difference between the populations.

Among our targets, the most straightforward to identify are the six HB stars. Being located in the red part of the HB (see Fig\,\ref{fig:bkg_fit}), they can be safely identified as 1G stars following the results of \citet[][which found all rHB in this cluster to have high {[}O/Fe{]} and low {[}Na/Fe{]}]{marino_2011}. This can be further reinforced by cross-matching our sample with the one in \citet{marino_2011}: we find two stars in common (W1068 and W2887) which have [Na/Fe]=$-0.09$ and $-0.06$, respectively. A value of [Na/Fe] close to zero indicates that the star belongs to the 1G \citep[see e.g.][]{Gratton2001,marino_2008,carretta_2009}, as 2G stars tend to have enhanced abundances of [Na/Fe] (and corresponding depleted abundances of [O/Fe]). This corroborates the identification of all rHB stars in our sample as 1G stars.

To separate the RGB stars between 1G and 2G we exploit the capability of the $\rm C_{U,B,I}=(U-B)-(B-I)$ index \citep{Milone_2013,Monelli_2013}. We summarize the procedure in Figure  \ref{fig:populations}. In a nutshell, we apply a Gaussian Mixture Model (GMM) to the verticalized RGB in the  $\rm C_{U,B,I}$ vs. V diagram (left and upper right panels of Figure \ref{fig:populations}). A free parameter for the Gaussian Mixture model is the number of clusters it finds. We tested the number of components across a range of 2 to 15 and based on an evaluation of the different Bayesian Information Criteria values that three clusters was the model that best described the data in this dimension. Our best-fit model has three components: two representing the main stellar populations in the cluster 1G (red) and 2G (blue) and one\footnote{with a peak at $\rm \Delta C_{U,B,I}\sim1.5$} (green) identifying the few binaries in our catalogue. This is in agreement with most literature sources that identify two stellar populations in this GC 
\citep[see][and references therein]{marino_2008,carretta_2009,carretta2013,marino_2011,milone_2017,marino_2017,lardo_2017,milone_2018,tailo_2019a}.

Our targets divide as follows: 10 1G stars and 16 2G stars. We took care to verify that the identification of our targets is not significantly affected by random factors. In order to assess whether the identification was affected wildly by the randomisation inert in the Gaussian Mixture Model computation and tested whether the identification was robust across 20 different initialization, finding that non of our targets switched identification. We highlight the position of our targets in the panels as large dots, color-coded as per population.

We get a further confirmation of the separation between the 1G and the 2G stars in our target sample by cross-matching our stars with the ones in \citet{marino_2008} and \cite{carretta_2009}. Only 12 stars are in common between these spectroscopic works and our sample and the spectroscopic and photometric identification agree very well, as reported in Table\ref{tab1}. One star (W2022) is classified as 1G in \citet{marino_2008} and 2G in \citet{carretta_2009}; we agree with the former, which is based on higher resolution spectra. We note that the ratio of 1G to 2G stars ($\sim40\%$ and $\sim60\%$, respectively) is compatible to the ratio found in the literature \cite{marino_2008,carretta_2009,carretta2013,marino_2011,milone_2017,marino_2017,lardo_2017}. 

\subsection{The average mass of the multiple populations}

\begin{table}[]
    \caption{Average mass for the 1G and 2G RGB stars in our sample.}
    \begin{adjustbox}{width=0.9\columnwidth}
    \centering
    \begin{tabular}{lccc}
        \hline
        \hline
                                            & 1G & 2G & 2G (excl. W4488)\\
        \hline
       $\rm \langle M_1/M_\odot \rangle$    &$0.893\pm0.043$ & $0.806\pm0.052$ &$0.897\pm0.017$\\ 
       $\rm \langle M_2/M_\odot \rangle$    &$0.822\pm0.021$ & $0.856\pm0.023$ &$0.838\pm0.011$\\ 
       $\rm \langle M_3/M_\odot \rangle$    &$0.859\pm0.014$ & $0.867\pm0.009$ &$0.868\pm0.009$\\ 
       $\rm \langle M_4/M_\odot \rangle$    &$0.887\pm0.033$ & $0.841\pm0.034$ &$0.886\pm0.014$\\ 
       \hline
    \end{tabular}
    \end{adjustbox}
    \label{tab3}
\end{table}

With our targets separated into first and second generation stars, we re-evaluate the average mass values. We find the average mass values reported in Table \ref{tab3}. We report the two separate sets of average mass values in the panels in Fig.\ref{fig:masses} as the red and blue solid lines, respectively for the 1G and the 2G, together with their $1\sigma$ interval. 
Considering our uncertainties on the derived masses, both statistical and systematical, our results do not allow to identify significant mass differences between the two populations (i.e. the mass difference is comparable with the errors). This is, however, unsurprising. 
Due to the low helium enhancement of the two stellar populations in M4 ($0.013$) and the low extension of the [Na/Fe]-[O/Fe], [(C+N)/Fe]-[O/Fe] and [Mg/Fe]-[Al/Fe] anti-correlations diagrams, compared to more massive and complex clusters like NGC\,2808 or M13, the difference between the evolving mass of the 1G and the 2G is predicted to be $\leq 0.017$ \citep{tailo_2019a}: a sensitivity hard to reach with this few targets in both populations when the masses are derived purely using the asteroseismic scaling relations instead of the more constraining individual mode frequencies.

\section{Evaluation of correction factor $f_{\Delta\nu}$}
\label{sec:sys_correction}

\begin{figure}
    \centering
    \includegraphics[width=\columnwidth]{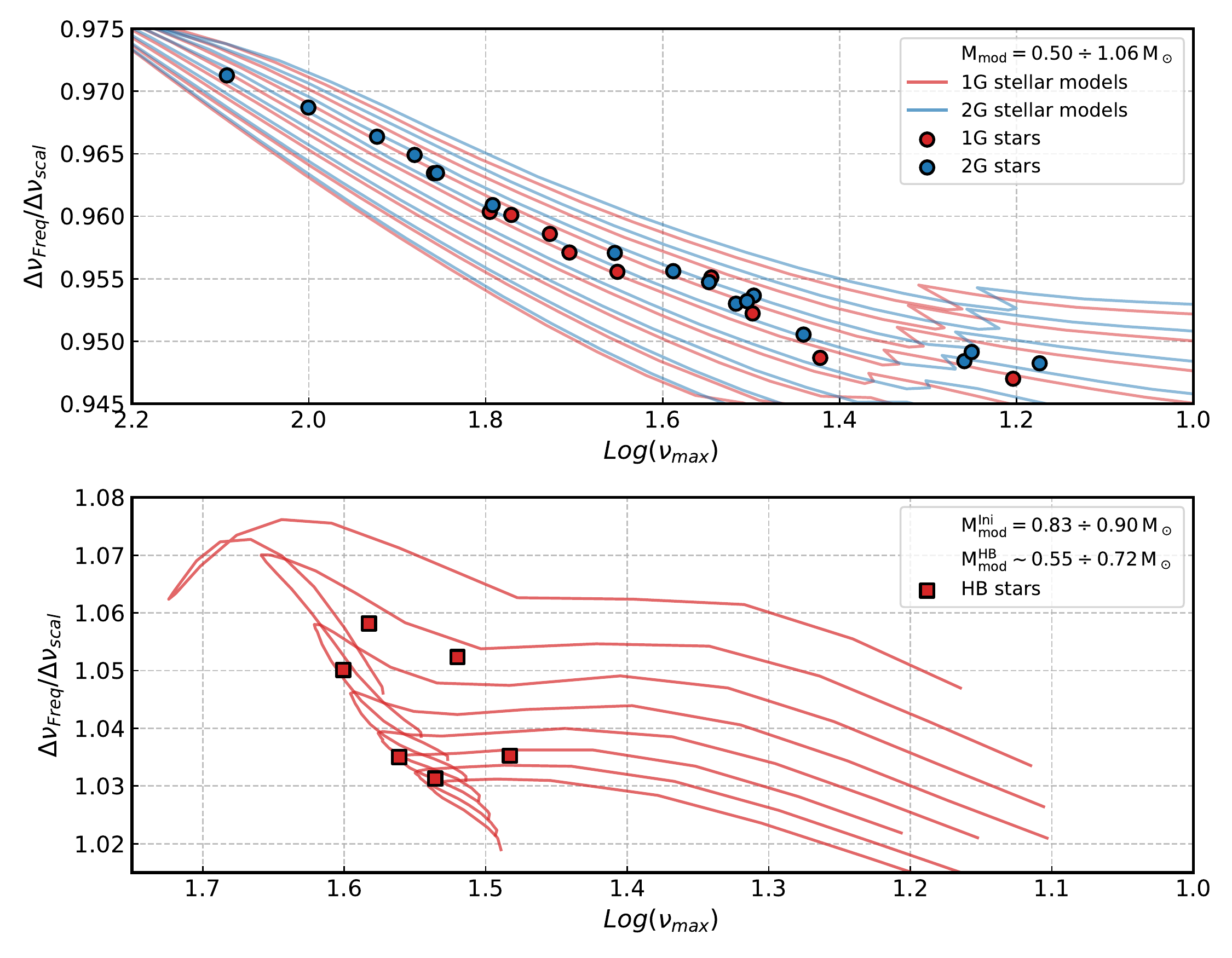}
    \caption{Correction values for our RGB targes (left), divided by populations, and our rHB stars (right). The solid lines are the corresponding models the correction are drawn from.}
    \label{fig:corr}
\end{figure}

Corrections to scaling relations linking $\Delta\nu$ to the mean density have been obtained with the iterative procedure described in the following.

For the RGB stars we calculate two grids of stellar models with the \textsc{mesa} software \citep{Paxton_2011,Paxton_2013,Paxton_2015,Paxton_2018,Paxton_2019}. Each grid is built with a Cartesian approach and spans the following dimensions as follows:: $\rm M$ from 0.50 to 1.20 $M_\odot$ with a step of 0.02$M_\odot$, [Fe/H]=$-1.1$, [$\alpha$/Fe]=+0.4 and standard helium (Y$\sim$0.25) for the 1G and $\rm M$ from 0.50 to 1.20 $M_\odot$ with a step of 0.02$M_\odot$, [Fe/H]=$-1.1$, [$\alpha$/Fe]=+0.3 and Y=0.263 for the 2G. The increased helium abundance and the lower [$\alpha$/Fe] is needed to simulate the different chemical abundance pattern observed in the 2G stars \citep{marino_2008,carretta_2009,carretta2013,tailo_2019a}. Each model is evolved up to the helium flash. The grid for the HB stars covers a range of initial masses that ranges from $\rm M=0.82\,M_\odot$ to $\rm 0.90\,M_\odot$ with a step of 0.01$M_\odot$ and the same [Fe/H] and [$\alpha$/Fe] of the 1G grid. We assumed a mass loss rate corresponding to a \citet{reimers_1975} parameter of $\rm \eta_R=0.47$, in line with the predictions of both \cite{tailo_2020} and \cite{mcdonald2015}. The HB models are evolved up to the end of the HB phase, including the helium flash, and arrive at the start of the HB phase with a mass in the $\rm M=0.56\,M_\odot$ to $\rm 0.72\,M_\odot$ range.

We then compute the individual oscillation frequencies for each model in the both grids using the \textsc{gyre} software \citep{Townsend_2013,Townsend_2018,Goldstein} to calculate individual oscillation frequencies. The correction to the $\Delta\nu$ for each track are then obtained with the method from \citet{Miglio_2016,Rodrigues_2017}. Once each single track has been built this way we use the mass calculated from Eq.3 and the $\nu_\mathrm{max}$ of each star to locate it along the track. For the RGB stars, the results of this iterative procedure are represented in the upper panel of Fig.\,\ref{fig:corr}. To calculate the correction to be applied to the rHB stars we locate our stars in the track corresponding to the HB mass obtained from Eq.3, and then locate the position of the tracks using the value of $\nu_\mathrm{max}$ as well. The results for the rHB stars is plotted in the lower panel of Fig.\,\ref{fig:corr}.

Compared with \cite{Miglio_2016} we used a larger range of correction values. Miglio and collaborators used a fixed value of $f_{\Delta\nu}=0.956$ which is fine for the limited range of $\rm log(\nu_{max})$ of their sample. In this works, changes to their procedure are necessary given the greater range of $\nu_{\rm max}$ and luminosities we are exploring. On the other hand, the values we adopted for the rHB stars are quite different from the $f_{\Delta\nu,HB}=1.029$ adopted by Miglio and collaborators. This is due to the lower mass loss they employ in their models. However, we find worth to note that models with lower mass loss -- resulting in more massive HB stars -- do not reach values of $\nu_\mathrm{max}$ compatible with the ones of some of our stars. The entire range of $f_{\Delta\nu}$ adopted is reported in Table \ref{tab2}.

\section{Full masses and radii sets}
\label{sec:dataset}

\begin{table*}[]
\caption{The complete data set of the masses and radii obtained in this work. Columns are: star id in \cite{Wallace19}, population identification according to the $C_{U,B,I}$ index, photometric radius ($\rm R_{CMD}$), uncorrected mass values from Eq.\,\ref{eq:scal1} to \ref{eq:scal4} ($\rm M^f_X$), uncorrected asteroseismic radius ($\rm R^f_S$), correction to the $\Delta\nu$, corrected mass values for Eq.\,\ref{eq:scal1}, \ref{eq:scal2} and \ref{eq:scal4} ($\rm M_X$), corrected asteroseismic radius ($\rm R_S$). Where appropriate each quantity is reported with its $1\sigma$ error. We assumed a flat 0.01 as the $1\sigma$ error for the correction.}
\begin{adjustbox}{width=2.0\columnwidth}
\centering
\renewcommand{\arraystretch}{1.2} 
\begin{tabular}{lllllllllllll}
\hline
\hline
  WID & POP$_{CUBI}$ &    $\rm R_{CMD}/R_\odot$ &     $\rm M^f_1/M_\odot$ &     $\rm M^f_2/M_\odot$ &     $\rm M_3/M_\odot$ &     $\rm M^f_4/M_\odot$ &      $\rm R^f_S/R_\odot$ &  $f_{\Delta\nu}$ & $\rm M_1/M_\odot$ & $\rm M_2/M_\odot$ & $\rm M_4/M_\odot$ &  $\rm R_S/R_\odot$ \\
\hline
W491 &  2G &  8.879$\pm$0.605 & 1.097$\pm$0.191 & 0.817$\pm$0.179 & 0.902$\pm$0.119 & 1.034$\pm$0.132 &  9.793$\pm$0.805 & 0.956 & 0.915$\pm$0.159 & 0.895$\pm$0.196 & 0.911$\pm$0.116 &  8.943$\pm$0.735 \\
 W508 &  2G &  6.298$\pm$0.406 & 1.175$\pm$0.180 & 0.793$\pm$0.161 & 0.904$\pm$0.114 & 1.086$\pm$0.125 &  7.179$\pm$0.491 & 0.965 & 1.018$\pm$0.156 & 0.852$\pm$0.173 & 0.983$\pm$0.113 &  6.684$\pm$0.457 \\
 W760 &  2G & 12.708$\pm$0.877 & 1.159$\pm$0.243 & 0.736$\pm$0.159 & 0.856$\pm$0.122 & 1.059$\pm$0.171 & 14.788$\pm$1.247 & 0.948 & 0.938$\pm$0.196 & 0.818$\pm$0.177 & 0.913$\pm$0.147 & 13.302$\pm$1.121 \\
 W779 &  1G &  7.431$\pm$0.503 & 0.993$\pm$0.169 & 0.825$\pm$0.174 & 0.877$\pm$0.119 & 0.956$\pm$0.125 &  7.906$\pm$0.559 & 0.959 & 0.838$\pm$0.143 & 0.897$\pm$0.189 & 0.850$\pm$0.111 &  7.264$\pm$0.513 \\
 W799 & rHB &  7.321$\pm$0.436 & 0.907$\pm$0.171 & 0.444$\pm$0.082 & 0.563$\pm$0.068 & 0.786$\pm$0.116 &  9.291$\pm$0.684 & 1.054 & 1.120$\pm$0.211 & 0.399$\pm$0.074 & 0.911$\pm$0.134 & 10.324$\pm$0.760 \\
 W837 &  2G &  4.591$\pm$0.284 & 0.930$\pm$0.128 & 0.734$\pm$0.142 & 0.794$\pm$0.096 & 0.887$\pm$0.092 &  4.967$\pm$0.300 & 0.971 & 0.827$\pm$0.114 & 0.778$\pm$0.150 & 0.817$\pm$0.085 &  4.686$\pm$0.283 \\
W1068 &  2G &  6.389$\pm$0.415 & 1.068$\pm$0.218 & 0.793$\pm$0.167 & 0.876$\pm$0.114 & 1.006$\pm$0.153 &  7.056$\pm$0.642 & 0.963 & 0.920$\pm$0.188 & 0.854$\pm$0.180 & 0.907$\pm$0.138 &  6.550$\pm$0.596 \\
W1091 &  2G &  9.921$\pm$0.673 & 1.160$\pm$0.396 & 0.813$\pm$0.207 & 0.916$\pm$0.128 & 1.081$\pm$0.267 & 11.169$\pm$1.791 & 0.954 & 0.960$\pm$0.327 & 0.894$\pm$0.227 & 0.946$\pm$0.234 & 10.158$\pm$1.629 \\
W1156 &  1G &  7.444$\pm$0.504 & 1.370$\pm$0.289 & 0.653$\pm$0.148 & 0.836$\pm$0.109 & 1.181$\pm$0.179 &  9.530$\pm$0.982 & 0.957 & 1.150$\pm$0.243 & 0.713$\pm$0.162 & 1.045$\pm$0.158 &  8.730$\pm$0.900 \\
W1225 & rHB &  7.862$\pm$0.483 & 0.516$\pm$0.160 & 0.742$\pm$0.171 & 0.657$\pm$0.080 & 0.555$\pm$0.125 &  6.964$\pm$1.016 & 1.031 & 0.583$\pm$0.181 & 0.698$\pm$0.160 & 0.605$\pm$0.136 &  7.406$\pm$1.080 \\
W1582 &  1G &  6.453$\pm$0.419 & 1.300$\pm$0.347 & 0.749$\pm$0.173 & 0.900$\pm$0.115 & 1.164$\pm$0.224 &  7.756$\pm$0.992 & 0.963 & 1.120$\pm$0.299 & 0.807$\pm$0.186 & 1.049$\pm$0.202 &  7.199$\pm$0.921 \\
W1608 & rHB &  8.228$\pm$0.498 & 0.775$\pm$0.181 & 0.587$\pm$0.111 & 0.644$\pm$0.082 & 0.733$\pm$0.135 &  9.026$\pm$0.781 & 1.035 & 0.890$\pm$0.208 & 0.548$\pm$0.103 & 0.807$\pm$0.148 &  9.673$\pm$0.837 \\
W1717 &  2G &  5.977$\pm$0.380 & 0.973$\pm$0.314 & 0.869$\pm$0.215 & 0.903$\pm$0.111 & 0.951$\pm$0.217 &  6.206$\pm$0.989 & 0.966 & 0.848$\pm$0.273 & 0.931$\pm$0.230 & 0.864$\pm$0.197 &  5.796$\pm$0.923 \\
W1763 &  2G & 13.993$\pm$0.971 & 1.142$\pm$0.387 & 0.734$\pm$0.161 & 0.851$\pm$0.142 & 1.046$\pm$0.278 & 16.214$\pm$2.021 & 0.948 & 0.923$\pm$0.313 & 0.816$\pm$0.180 & 0.901$\pm$0.240 & 14.579$\pm$1.818 \\
W1912 &  2G &  5.336$\pm$0.336 & 0.918$\pm$0.469 & 0.836$\pm$0.263 & 0.863$\pm$0.106 & 0.901$\pm$0.324 &  5.506$\pm$1.397 & 0.969 & 0.809$\pm$0.413 & 0.891$\pm$0.281 & 0.825$\pm$0.296 &  5.167$\pm$1.311 \\
W2021 &  1G &  7.957$\pm$0.539 & 0.942$\pm$0.185 & 0.797$\pm$0.176 & 0.843$\pm$0.112 & 0.911$\pm$0.131 &  8.413$\pm$0.768 & 0.956 & 0.785$\pm$0.154 & 0.873$\pm$0.192 & 0.802$\pm$0.116 &  7.682$\pm$0.701 \\
W2022 &  1G & 13.220$\pm$0.911 & 0.996$\pm$0.194 & 0.740$\pm$0.162 & 0.817$\pm$0.113 & 0.939$\pm$0.139 & 14.594$\pm$1.216 & 0.947 & 0.801$\pm$0.156 & 0.826$\pm$0.180 & 0.806$\pm$0.119 & 13.089$\pm$1.091 \\
W2034 &  1G & 10.139$\pm$0.692 & 1.070$\pm$0.304 & 0.692$\pm$0.163 & 0.800$\pm$0.112 & 0.981$\pm$0.205 & 11.726$\pm$1.509 & 0.949 & 0.867$\pm$0.246 & 0.769$\pm$0.181 & 0.846$\pm$0.177 & 10.553$\pm$1.358 \\
W2162 &  2G &  9.341$\pm$0.636 & 1.160$\pm$0.429 & 0.805$\pm$0.213 & 0.909$\pm$0.126 & 1.079$\pm$0.287 & 10.553$\pm$1.859 & 0.955 & 0.964$\pm$0.356 & 0.883$\pm$0.234 & 0.947$\pm$0.252 &  9.619$\pm$1.695 \\
W2386 & rHB &  7.891$\pm$0.485 & 0.648$\pm$0.106 & 0.730$\pm$0.143 & 0.701$\pm$0.083 & 0.663$\pm$0.080 &  7.582$\pm$0.564 & 1.035 & 0.743$\pm$0.121 & 0.681$\pm$0.134 & 0.730$\pm$0.088 &  8.122$\pm$0.604 \\
W2665 &  2G & 12.783$\pm$0.879 & 1.163$\pm$0.186 & 0.730$\pm$0.159 & 0.853$\pm$0.114 & 1.059$\pm$0.124 & 14.927$\pm$1.117 & 0.949 & 0.943$\pm$0.151 & 0.811$\pm$0.177 & 0.915$\pm$0.108 & 13.447$\pm$1.006 \\
W2678 &  1G &  9.716$\pm$0.662 & 1.203$\pm$0.388 & 0.750$\pm$0.192 & 0.878$\pm$0.117 & 1.095$\pm$0.251 & 11.377$\pm$1.796 & 0.952 & 0.989$\pm$0.319 & 0.827$\pm$0.212 & 0.955$\pm$0.219 & 10.315$\pm$1.629 \\
W2887 & rHB &  6.969$\pm$0.415 & 2.803$\pm$0.791 & 0.288$\pm$0.059 & 0.614$\pm$0.080 & 1.778$\pm$0.377 & 14.885$\pm$1.811 & 1.052 & 3.437$\pm$0.970 & 0.260$\pm$0.054 & 2.051$\pm$0.435 & 16.483$\pm$2.006 \\
W3033 &  2G &  9.414$\pm$0.638 & 1.097$\pm$0.393 & 0.769$\pm$0.200 & 0.866$\pm$0.120 & 1.022$\pm$0.264 & 10.597$\pm$1.803 & 0.953 & 0.905$\pm$0.324 & 0.847$\pm$0.220 & 0.893$\pm$0.231 &  9.624$\pm$1.638 \\
W3041 &  1G &  6.758$\pm$0.441 & 0.985$\pm$0.512 & 0.793$\pm$0.255 & 0.852$\pm$0.110 & 0.943$\pm$0.346 &  7.266$\pm$1.867 & 0.960 & 0.838$\pm$0.436 & 0.859$\pm$0.276 & 0.842$\pm$0.309 &  6.701$\pm$1.722 \\
W3073 &  2G &  6.809$\pm$0.460 & 0.942$\pm$0.213 & 0.817$\pm$0.188 & 0.857$\pm$0.112 & 0.916$\pm$0.147 &  7.140$\pm$0.790 & 0.961 & 0.803$\pm$0.181 & 0.885$\pm$0.204 & 0.819$\pm$0.132 &  6.593$\pm$0.729 \\
W3480 &  2G &  8.097$\pm$0.529 & 1.092$\pm$0.187 & 0.795$\pm$0.165 & 0.884$\pm$0.114 & 1.025$\pm$0.131 &  9.001$\pm$0.695 & 0.957 & 0.916$\pm$0.157 & 0.868$\pm$0.180 & 0.906$\pm$0.116 &  8.244$\pm$0.637 \\
W3528 &  1G &  7.066$\pm$0.458 & 1.141$\pm$0.234 & 0.777$\pm$0.158 & 0.883$\pm$0.119 & 1.056$\pm$0.167 &  8.030$\pm$0.661 & 0.960 & 0.969$\pm$0.199 & 0.843$\pm$0.172 & 0.943$\pm$0.149 &  7.402$\pm$0.609 \\
W3564 &  2G &  9.664$\pm$0.654 & 1.086$\pm$0.288 & 0.803$\pm$0.187 & 0.888$\pm$0.121 & 1.022$\pm$0.198 & 10.688$\pm$1.312 & 0.953 & 0.896$\pm$0.237 & 0.883$\pm$0.206 & 0.894$\pm$0.173 &  9.712$\pm$1.192 \\
W3742 &  1G &  9.647$\pm$0.661 & 1.258$\pm$0.379 & 0.838$\pm$0.206 & 0.959$\pm$0.132 & 1.159$\pm$0.252 & 11.047$\pm$1.573 & 0.955 & 1.047$\pm$0.315 & 0.918$\pm$0.226 & 1.019$\pm$0.222 & 10.077$\pm$1.435 \\
W3929 & rHB &  7.067$\pm$0.422 & 1.100$\pm$0.318 & 0.448$\pm$0.097 & 0.604$\pm$0.070 & 0.919$\pm$0.194 &  9.535$\pm$1.271 & 1.058 & 1.379$\pm$0.398 & 0.400$\pm$0.087 & 1.076$\pm$0.227 & 10.677$\pm$1.423 \\
W4488 &  2G & 10.131$\pm$0.690 & 0.515$\pm$0.134 & 1.065$\pm$0.229 & 0.836$\pm$0.125 & 0.595$\pm$0.121 &  7.950$\pm$0.798 & 0.951 & 0.420$\pm$0.110 & 1.179$\pm$0.254 & 0.517$\pm$0.105 &  7.183$\pm$0.721 \\
\hline
\hline
\end{tabular}
\end{adjustbox}
\label{tab2}
\end{table*}

We report in Table \ref{tab2} the complete set of obtained masses, together with the photometric and asteroseismic radii for our sample of stars. Two sets of values are reported, respectively with uncorrected and corrected $\Delta\nu$. 

\section{Comparison of the radii}
\label{sec:sanity_check}

\begin{figure}
    \centering
    \includegraphics[width=\columnwidth]{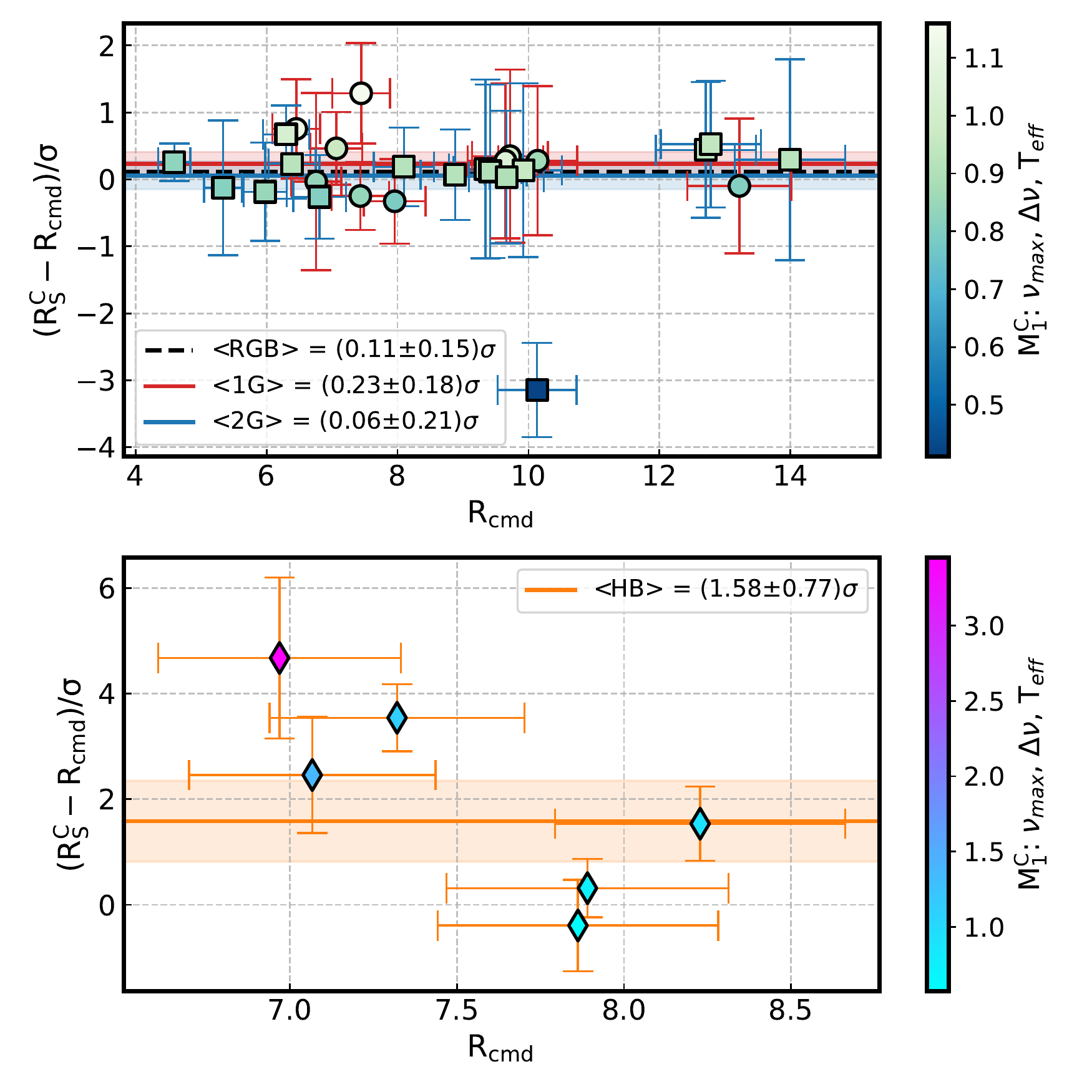}
    \caption{Difference between the photometric ($R_\mathrm{CMD}$) and the asteroseismic ($R_\mathrm{S}$) radii for our RGB (top) and rHB (bottom) stars. The black, dashed line in the upper panel is the average dispersion of the entire sample, while the coloured lines represent the average value of each populations with their 1$\sigma$ intervals as the shaded areas}
    \label{fig:radii}
\end{figure}

As shown in \S\,\ref{sec:dnu} and in Fig.\,\ref{fig:dnu_test}, the value of $\Delta\nu$ can vary wildly, especially for the HB stars. This causes the masses and radii of the stars in our sample to vary as well, leading to severely underestimating or overestimating both. 

To conduct a simple but effective quality check of our results we compare the radius obtained from the asteroseismic parameters ($R_\mathrm{S}$) with the one obtained from the photometry ($R_\mathrm{CMD}$). 
The results are plotted in Fig.\,\ref{fig:radii}, respectively for the RGB (top) and the HB stars (bottom). We see that, after the correction on $\Delta\nu$ is applied, the radius of most of our RGB targets fall within $|\langle(R_\mathrm{S}-R_\mathrm{CMD})/\sigma\rangle|<1$. The only exception is one 2G stars (W4488) that is located at $\langle(R_\mathrm{S}-R_\mathrm{CMD})/\sigma\rangle \sim -3\sigma$. Since the mass obtained from Eq.3 is $0.836\pm0.072\,M_\odot$, in line with the average of its 2G counterparts, we have strong hints that the culprit of this large discrepancy is a wrong measurement of $\Delta\nu$. If we remove it from the sample, then the average mass of our RGB stars becomes $0.879\pm0.025 M_\odot$, $0.836\pm0.012 M_\odot$, $0.862\pm0.008 M_\odot$ and $0.882\pm0.026 M_\odot$, respectively for Eq.\ref{eq:scal1} to \ref{eq:scal4}.

The results of this check for the rHB stars in our sample yields a different result however.  Because of the difficulties associated to measuring $\Delta\nu$, the radii of the rHB stars varies up to the $|\langle(R_\mathrm{S}-R_\mathrm{CMD})/\sigma\rangle|<5$ level, with only two (namely W1225 and W2386) falling in the 1~$\sigma$ range. The points in the bottom panel of Fig.\ref{fig:radii} are also colour-coded according the value of $M_1$, showing that when the radius is overestimated the mass follows as well (and vice versa). 

We make an additional test by searching for the value of $f_{\Delta\nu}$ that gives us $|\langle(R_\mathrm{S}-R_\mathrm{CMD})/\sigma\rangle|<1$ and average masses from Eq.\ref{eq:scal1} to \ref{eq:scal4} in agreement with each other for our 6 rHB stars. We found that with $f_{\Delta\nu}\sim0.940$ these two conditions are reasonably satisfied. Indeed, we obtained $\langle(R_\mathrm{S}-R_\mathrm{CMD})/\sigma\rangle =-0.19\pm0.76$ and a set of average masses of $0.569\pm0.298 M_\odot$, $0.512\pm0.096 M_\odot$, $0.632\pm0.024 M_\odot$ and $0.606\pm0.168 M_\odot$. However the correction needed is not compatible with the one predicted from the model plotted in Fig.\,\ref{fig:corr} and in other literature studies \citep{Miglio_2013,Rodrigues_2017} for low-mass helium-burning stars. Finally, if we impose $R_\mathrm{CMD}=R_\mathrm{S}$ we obtain values for $f_{\Delta\nu}$ that vary wildly and that are much different from the ones that can be obtained from the models in Fig.\ref{fig:corr}. We can therefore exclude any major role of the correction to $\Delta\nu$ in the large spread in $\langle(R_\mathrm{S}-R_\mathrm{CMD})/\sigma\rangle$ observed among our rHB stars. 

Based on these considerations, we only use the results following Eq. 3 from HB stars as they are not dependent on $\Delta\nu$. Regardless,  the mass estimates obtained with all four equations are reported in Table\ref{tab2} for completeness.

\section{Systematics}
\label{sec:syst_all}

\begin{figure*}
    \centering
    \includegraphics[width=1.90\columnwidth]{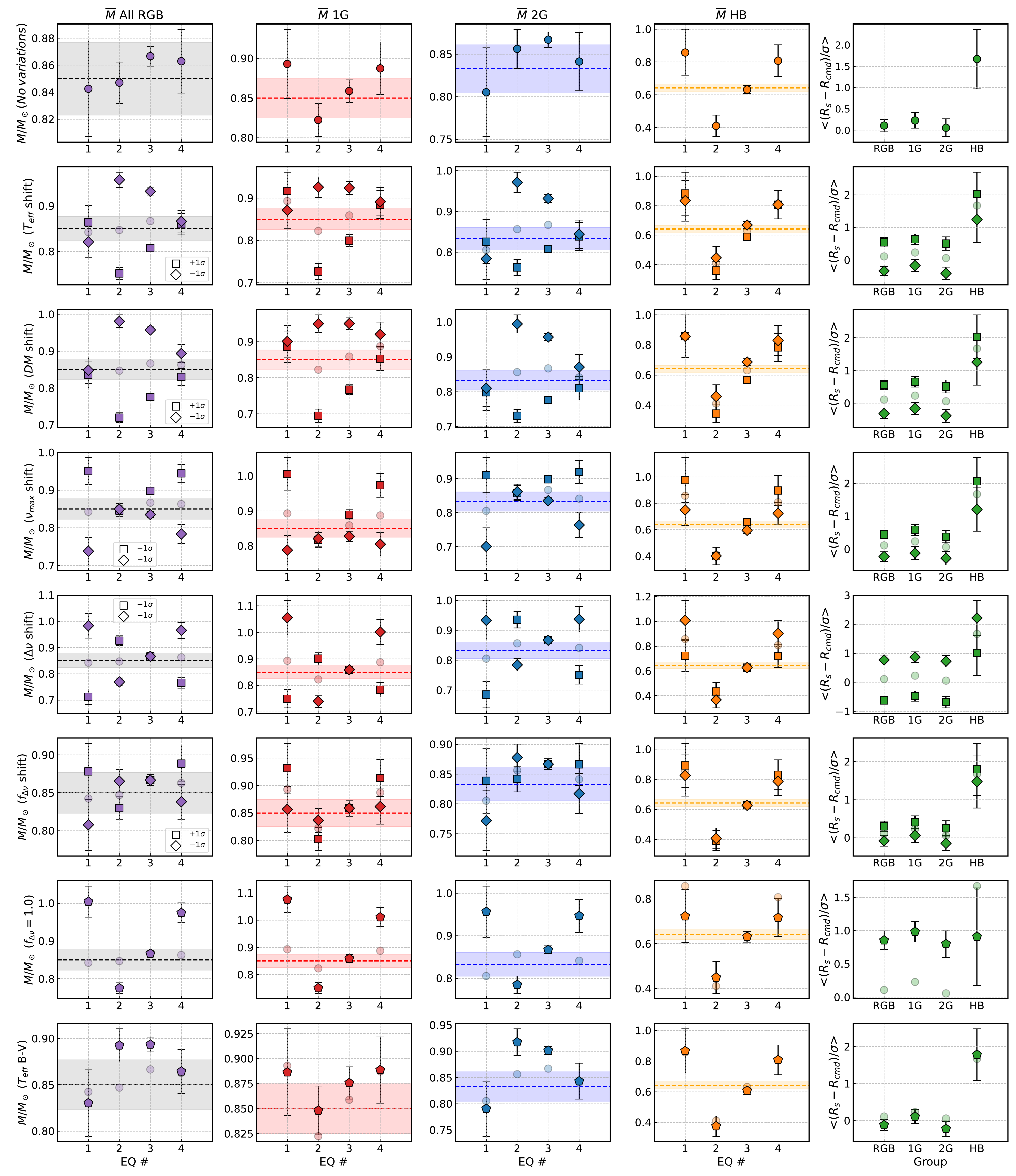}
    \caption{Systematic shift of the average mass and radii dispersion for different groups of stars. The first row reports the non shifted values, which are also reported as transparent points in each panel. Each row past the first explore the effects of a shift in a given parameter of its $1~\sigma$ interval, as labelled. The last two rows report the non-corrected mass and dispersion average values, and what happens when the Temperatures coming from the $B-V$ bands are used. The dashed lines represent the mass derived from the EB results in \citet[][0.85$M_\odot$,first column]{Kaluzny2013}, the mass of the 1G, the 2G and the HB from \citet[][the values are 0.85$M_\odot$, 0.833$M_\odot$, 0.64$M_\odot$, respectively for the second to fourth columns]{tailo_2019a} and the shaded area mark their 1$\sigma$  intervals.}
    \label{fig:syst_all}
\end{figure*}

We discuss several systematical effects that can affect our results in this subsection. The results of this exploration are collectively reported in Figure \ref{fig:syst_all}. Each row starting from the second explores a different systematical shift and its effect on the average mass determination (first four columns) and on $\langle(R_\mathrm{S}-R_\mathrm{CMD})/\sigma\rangle$ (last column) of the individual groups of stars in our sample\footnote{The entirety of the RGB stars, the 1G, the 2G and the HB stars}. The first row reports the non-shifted mass and scatter values that are also reported in each panel as the transparent points for clarity and ease of comparison. In each panel describing a mass shift we report independent estimates coming from different sources as the coloured dashed line: the extrapolated mass from the eclipsing binaries (EB) in \citet[][0.85$M_\odot$, first column]{Kaluzny2013} and the mass of the 1G (red), the 2G (blue) and the rHB stars (orange) from \citet[][values are 0.85$M_\odot$, 0.833$M_\odot$, 0.64$M_\odot$, respectively for the second to fourth columns]{tailo_2019a} with their respective $1\sigma$ interval.

We explore what happens to our mass and $\langle(R_{\rm S}-R_{\rm CMD})/\sigma\rangle$ estimates when the four main parameters in the scaling relations are shifted by $\pm1\sigma$. We report the results in the second to fourth rows of Fig.\,\ref{fig:syst_all} respectively for a shift in $T_{\rm eff}$, distance modulus (hence luminosity), $\nu_{\rm max}$ and $\Delta\nu$. The variation of the average mass is proportional to the power that the specific parameters appears in the scaling equations. The intensity of the shift is however proportional to the power that specific parameters appears in the specific scaling equation. This means that e.g. the mass shift in $\Delta\nu$ is more pronounced for Eq.1 than Eq.2, and so on. Similar considerations can be drawn for $\langle(R_{\rm S}-R_{\rm CMD})/\sigma\rangle$, where the shift for the RGB radii is significant but does not go past the $|\langle(R_{\rm S}-R_{\rm CMD})/\sigma\rangle|>1$ limit. 

The values of the correction to $\Delta \nu$ can be somewhat model-specific for several reasons connected to the exact physics used to calculate the underlying stellar models.
To explore the systematic introduced by the procedure to obtain the correction, we shift the values of $f_{\Delta\nu}$ reported in Table \ref{tab2} by $\pm0.01$. The results are reported in the third-to-last row of Fig.\,\ref{fig:syst_all}. We see that, while a shift is present in all groups in the majority of cases it is within $1\sigma$. The same can be said for the shift in  $\langle(R_\mathrm{S}-R_\mathrm{CMD})/\sigma\rangle$ in the last column of the same row. For completeness in the second-to-last row of the figure we explore what happens when corrections are not taken into account (i.e. $f_{\Delta\nu}=1.0$ for all stars).   

\begin{figure}
    \centering
    \includegraphics[width=\columnwidth]{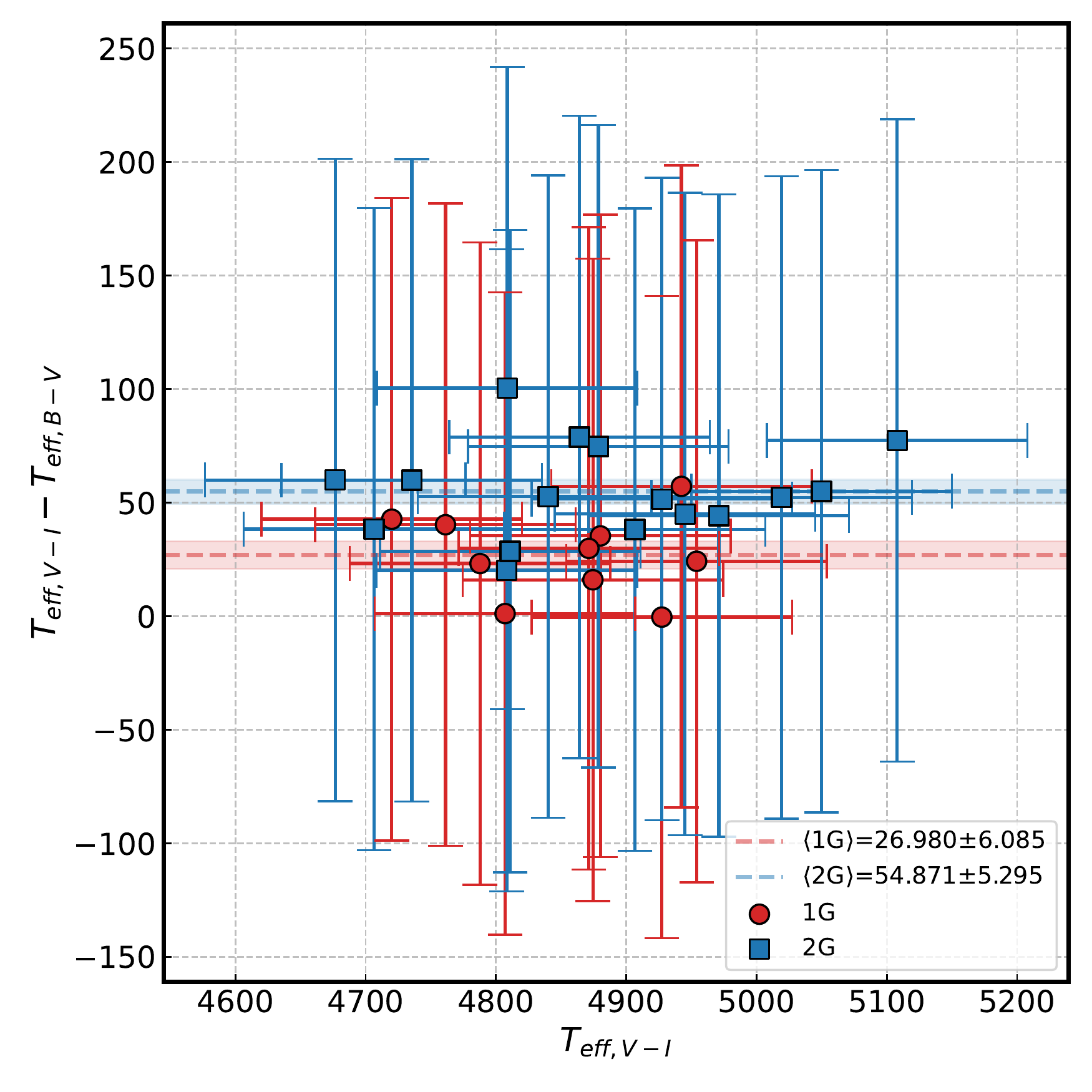}
    \caption{Shift in temperatures for our RGB targets when the estimate is done with the $B-V$ or the $V-I$ colours. The dashed lines are the average values for the two populations with their respective $1 \sigma$ interval.}
    \label{fig:comp_temp}
\end{figure}

When studying an old GC, which is made up by multiple stellar populations, differing in light elements abundances, it is important to check whether the bands used in the temperature evaluation are affected by the star-to-star variation of the chemical patterns. To show this we compare the temperatures estimates derived from the $B-V$ and $V-I$ colours. The results of this comparison are plotted in Fig.\,\ref{fig:comp_temp}. We see that the star to star variation of the temperature is well within the 1$\sigma$ range, however the shift in average temperature of the two stellar populations, while small -- $\rm \sim 27$~K for the 1G and $\rm \sim 55$~K for the 2G -- is significant and affects, collectively, the 2G stars more. For the 2G stars, this shift occur as the $B$ band is more sensitive to the changes in [C/Fe] and [N/Fe]. This small temperature difference is enough to alter the average mass difference of the two populations.

 We show what happens to our average mass estimates when we adopt the temperatures coming from the $B-V$ colours in the bottom row of Fig.\ref{fig:syst_all}. We see that the agreement with the independent mass estimate for M4 gets slightly worse, because the lower temperature values produce slightly higher mass estimates. The shift is greater for the 2G RGB stars while within the 1~$\sigma$ range for the 1G RGB and rHB stars. Because the 2G stars are the majority they also affect the agreement of the combined sample. Similar considerations hold for the radii dispersion, shown in the leftmost panel of the bottom row in the Figure.

In the case of M4, studied in this work, the variations are minimal, but in more complex and interesting cases, like 47~Tuc \citep[see e.g.][for a showcase]{carretta2013,milone_2017}, these effects can be relevant to the results. We then conclude that the best results are achieved with combination of bands less affected by the abundance pattern shifts in the multiple populations.

\end{appendix}
\end{document}